\documentclass[sigplan,10pt]{acmart}
\usepackage{booktabs}
\usepackage{graphicx}
\usepackage{textcomp}
\usepackage[table]{xcolor}
\usepackage{multirow}
\usepackage{subcaption}
\usepackage{pifont}
\usepackage{hyperref} 
\usepackage{enumitem}

\usepackage{booktabs}
\usepackage{multirow}
\usepackage{makecell}

\usepackage{algorithm}
\usepackage{algpseudocode}
\usepackage{amsmath}

\usepackage{xspace}

\usepackage[textsize=small]{todonotes}

\usepackage{url}

\usepackage[normalem]{ulem} 

\newcommand{\sysname}{OLED-MoE\xspace}

\newcommand\paraspace{\vspace*{0.25ex}}
\providecommand\parab[1]{\paraspace\noindent\textbf{#1}}

\newcommand{\secref}[1]{\S\ref{#1}}
\newcommand{\figref}[1]{Figure~\ref{#1}}

\newif\ifshowrevise
\showrevisetrue  

\newcommand{\revise}[2]{%
  \ifshowrevise
    \ifx&#1&\else{\color[RGB]{0,0,192}\sout{#1}}\fi  
    \ifx&#1&\else\ \fi  
    \ifx&#2&\else{\color[RGB]{192,0,0}{#2}}\fi  
  \else#2\fi 
}

\newif\ifshowtodos
\showtodostrue  

\AtBeginDocument{%
  }

\copyrightyear{2027}
\acmYear{2027}
\setcopyright{cc}
\setcctype{by}
\acmConference[EuroSys '27]{22nd European Conference on Computer Systems}{April 19--23, 2027}{Rabat, Morocco}
\acmBooktitle{22nd European Conference on Computer Systems (EuroSys '27), April 19--23, 2027, Rabat, Morocco}
\acmDOI{10.1145/3842654.3848535}
\acmISBN{979-8-4007-2971-3/2027/04}

\begin{document}
\title[OLED-MoE: Accelerating dLLM via Locality-Aware Expert Offloading]{OLED-MoE: Accelerating MoE-Based dLLM Inference via Inter-Iteration Locality-Aware Expert Offloading}

\author{Jingyuan Xiao}
\orcid{0009-0008-1753-4144}
\affiliation{%
  \institution{Tianjin University}
  \city{Tianjin}
  \country{China}}
\email{3020244304@tju.edu.cn}

\author{Jiayue Wang}
\orcid{0009-0003-7622-9928}
\affiliation{%
  \institution{Tianjin University}
  \city{Tianjin}
  \country{China}}
\email{wangjiayue05@tju.edu.cn}

\author{Yitao Hu}
\orcid{0009-0004-0458-0900}
\authornote{Corresponding author.}
\affiliation{%
  \institution{Tianjin University}
  \city{Tianjin}
  \country{China}}
\email{yitao@tju.edu.cn}

\author{Xinning Wang}
\orcid{0009-0004-7239-5862}
\affiliation{%
  \institution{Tianjin University}
  \city{Tianjin}
  \country{China}}
\email{wxn\_coic@tju.edu.cn}

\author{Shi Chen}
\orcid{0009-0003-5267-8649}
\affiliation{%
  \institution{Tianjin University}
  \city{Tianjin}
  \country{China}}
\email{jiangnan9@tju.edu.cn}

\author{Ziqi Gong}
\orcid{0009-0002-1699-8592}
\affiliation{%
  \institution{Tianjin University}
  \city{Tianjin}
  \country{China}}
\email{victayria@hotmail.com}

\author{Zhengchao Wang}
\orcid{0009-0005-0480-0448}
\affiliation{%
  \institution{Tianjin University}
  \city{Tianjin}
  \country{China}}
\email{right135@tju.edu.cn}

\author{Guotao Yang}
\orcid{0009-0005-9355-9296}
\affiliation{%
  \institution{Tianjin University}
  \city{Tianjin}
  \country{China}}
\email{gtyang@tju.edu.cn}

\author{Sheng Chen}
\orcid{0000-0001-7038-4407}
\affiliation{%
  \institution{Tianjin University}
  \city{Tianjin}
  \country{China}}
\email{chensheng@tju.edu.cn}

\author{Keqiu Li}
\orcid{0000-0003-3089-7907}
\affiliation{%
  \institution{Tianjin University}
  \city{Tianjin}
  \country{China}}
\email{keqiu@tju.edu.cn}

\renewcommand{\shortauthors}{Jingyuan Xiao et al.}

\begin{abstract}
Semi-autoregressive diffusion large language models (dLLMs) improve decoding parallelism through iterative block-wise denoising, but scaling them with mixture-of-experts (MoE) layers introduces a large expert parameter footprint that exceeds memory-constrained GPU capacity. Expert offloading is a natural remedy, yet existing MoE serving systems target autoregressive decoding and rely on intra-iteration layer-wise prefetching: while computing one layer, they predict and load experts for subsequent layers. Under dLLM inference, block-wise routing expands the active expert working set within each iteration, making such prefetches difficult to complete in time and costly when mispredicted. Consequently, existing prefetch-based solutions often degenerate into on-demand expert loading with high decoding latency.

We propose \sysname, an expert offloading system that shifts the optimization target from intra-iteration prefetching to inter-iteration expert retention. Its key insight is that adjacent denoising iterations exhibit strong expert routing overlap, and token confidence indicates which experts are likely to be reused. \sysname uses confidence-guided inter-iteration prediction to retain high-value experts in GPU memory without introducing extra prefetch traffic. It further compensates unavoidable cache misses through CPU-GPU cooperative execution, jointly considering dynamic expert computation load and predicted future reuse. Across diverse dLLM workloads, \sysname reduces time per output token (TPOT) by 1.23$\times$--7.93$\times$ and improves expert cache utilization by 1.44$\times$--4.23$\times$ over state-of-the-art offloading systems. Notably, \sysname approaches full-residency performance while using only 40\% of the expert GPU memory, incurring merely 23\% higher TPOT despite a 60\% reduction in expert memory footprint. OLED-MoE's source code is publicly available at
https://github.com/flashserve/OLED-MoE.

\end{abstract}

\ccsdesc[500]{Computing methodologies~Machine learning}
\ccsdesc[500]{Computer systems organization~Architectures}

\keywords{Diffusion Large Language Models; LLM Inference; Mixture-of-experts; Expert Offloading}

\maketitle
\section{Introduction}

\begin{figure}[t]
    \centering
    \includegraphics[width=0.9\columnwidth]{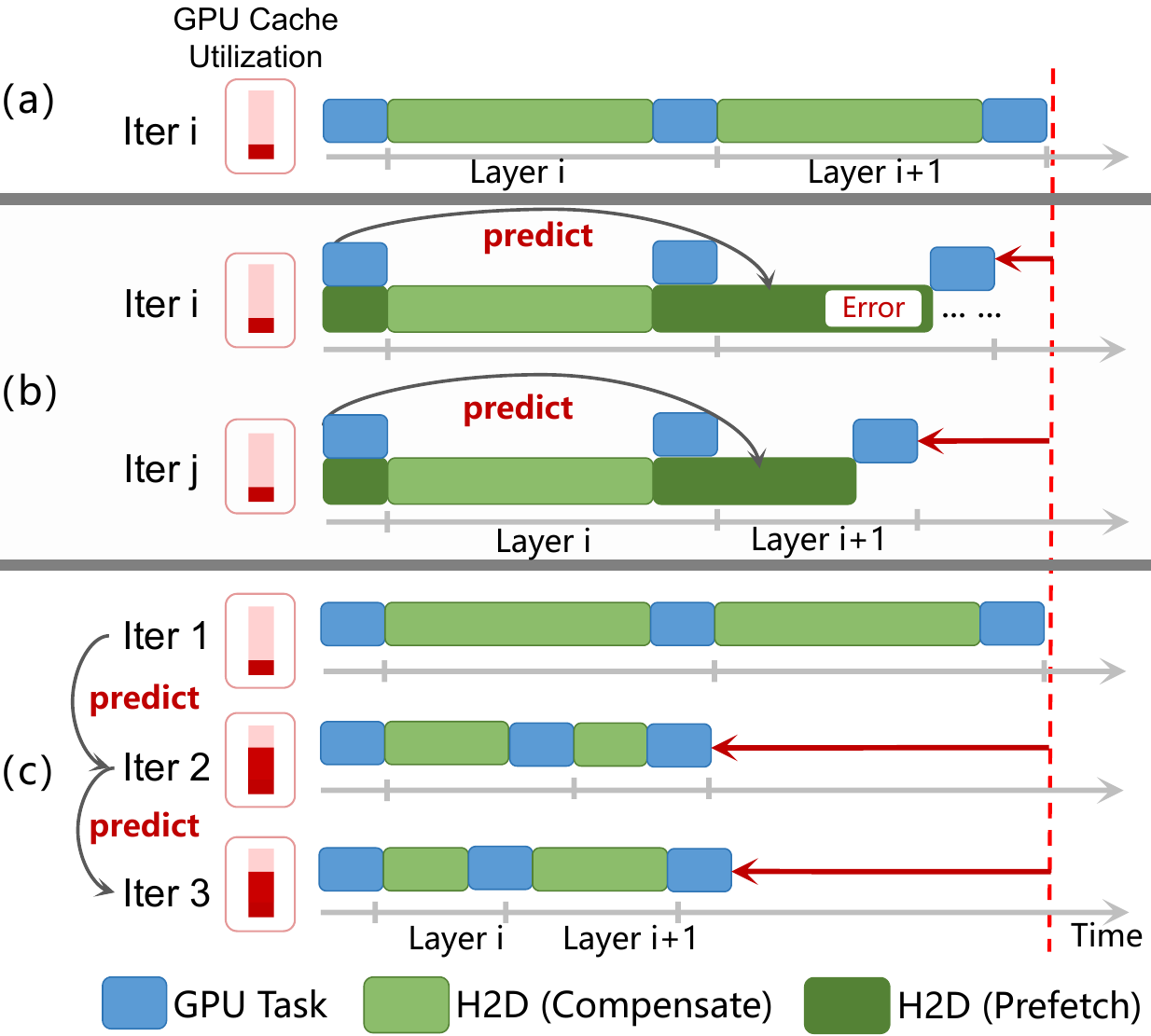}
    \caption{Expert offloading paradigms: (a) on-demand compensation, (b) intra-iteration prefetching, and (c) inter-iteration retention.}
    \label{fig:paradigm_compare}
\end{figure}

Large language models (LLMs) have traditionally relied on autoregressive (AR) decoding, which generates tokens strictly from left to right~\cite{xiong2024autoregressive,fu2025efficient,kim2025beyond,xie2024towards,wang2024loong}. Diffusion-based LLMs (dLLMs), such as LLaDA, have recently emerged as an alternative by formulating generation as iterative denoising and enabling semi-autoregressive block-wise decoding~\cite{tseng2025diffusion,yu2025discrete,li2025survey,lin2026efficient,kim2025cdlm}. To scale model capacity, recent dLLMs further adopt mixture-of-experts (MoE) layers~\cite{wei2026team,chen2026dynamic,lladamoe}, whose large expert footprint makes full GPU residency difficult on memory-constrained devices. Expert offloading addresses this by caching only a subset of experts in GPU memory, while placing the rest in host memory for GPU transfer or CPU execution when activated.

Existing MoE serving systems~\cite{ktransformers,fiddler,moeinfinity,swapmoe,sida,finemoe} are primarily designed around AR decoding. In AR generation, each decoding iteration generates one token, producing a relatively small active expert set at each layer. This leaves enough intra-iteration slack for layer-wise prefetching: while computing the current layer, the system predicts and transfers experts needed by subsequent layers. Remaining misses are compensated through H2D transfers or CPU-side expert computation. These mechanisms work only when prefetching and compensation can be hidden behind ongoing GPU computation.

This latency-hiding premise breaks under MoE-based dLLM inference. In each denoising iteration, a dLLM routes a block of tokens through MoE layers, expanding the active expert working set by up to about $8\times$ compared with AR generation (\secref{subsec:3_1}). The larger working set creates two coupled bottlenecks: more experts must be fetched or executed within the same iteration, while delayed or mispredicted prefetches consume PCIe bandwidth and GPU cache space needed to compensate current misses. As a result, intra-iteration prefetching often degenerates toward on-demand expert compensation, where H2D transfers and CPU compensation fall onto the critical path of decoding. \figref{fig:paradigm_compare} contrasts these paradigms. Under this regime, offloading performance is dominated by how effectively the limited GPU expert cache avoids repeated misses.

This cache-centric bottleneck motivates us to reframe dLLM expert scheduling from intra-iteration prefetching to inter-iteration expert retention. The opportunity comes from dLLM denoising itself: multiple iterations refine the same token block, and their same-layer expert sets strongly overlap across adjacent iterations (\secref{subsec:3_3}). Moreover, high-confidence tokens tend to preserve more stable routes, making their activated experts more likely to be reused in the next iteration (\secref{subsec:4_3}). This confidence signal is already produced by dLLM denoising. Instead of launching additional prefetches, it allows the system to rank currently resident experts by their near-future reuse value. This makes it possible to preserve high-value resident experts for future reuse rather than repeatedly fetching large expert sets within an iteration.

Based on this observation, we present \sysname, an expert offloading system tailored to MoE-based dLLMs. The key challenge is to make inter-iteration locality actionable for cache retention and miss compensation under limited GPU memory.

First, the system must translate set-level expert overlap into precise cache-retention decisions. Inter-iteration locality indicates that many experts are likely to be reused, but not which experts should be preserved under a limited GPU cache budget. A naive policy that retains recently activated experts can keep transient experts selected by unstable tokens, while evicting experts more likely to reappear in the next denoising iteration. 
This is further complicated by a granularity mismatch: dLLM confidence is produced at the token level, whereas cache management retains or evicts whole experts under layer-specific cache budgets. Meanwhile, heterogeneous activation and reuse patterns across layers make uniform cache allocation inefficient. \sysname addresses this challenge by mapping token confidence and gate scores into expert-level retention priorities and using layer-adaptive cache allocation to preserve the most valuable working set for the next denoising iteration.

Second, the system must compensate unavoidable cache misses without sacrificing future cache utility. Even with accurate reuse prediction, memory-constrained GPUs cannot retain all potentially activated experts, so missed experts must be either transferred to the GPU or executed on the CPU. A current-latency-only policy is insufficient: executing a high-reuse expert on the CPU may reduce current H2D traffic but cause repeated future misses, while transferring low-reuse experts may waste PCIe bandwidth and pollute the limited GPU cache. \sysname addresses this trade-off with a dynamic load- and reuse-aware cooperative compensation engine that jointly considers dynamic expert load and predicted future reuse, so compensation both resolves current misses and warms up the GPU cache for later iterations.

In summary, we make the following contributions:
\begin{itemize}
    \item We identify the mismatch between AR-oriented MoE offloading and MoE-based dLLM inference, showing that block-wise routing makes intra-iteration expert prefetching ineffective under dLLM workloads.

    \item We reveal inter-iteration expert locality in dLLMs and identify token confidence as a lightweight signal for expert reuse prediction.

    \item We design \sysname, an expert offloading system that leverages inter-iteration locality for confidence-guided retention, layer-adaptive cache allocation, and load- and reuse-aware cooperative compensation.

    \item We implement and evaluate \sysname across diverse dLLM workloads, reducing TPOT by 1.23$\times$--7.93$\times$ and improving expert cache utilization by 1.44$\times$--4.23$\times$ over state-of-the-art offloading systems.
\end{itemize}

\section{Background}\label{sec:2}

\subsection{Semi-Autoregressive MoE-based dLLMs}\label{subsec:2_1}

\begin{figure}[t]
    \centering
    \includegraphics[width=0.88\columnwidth]{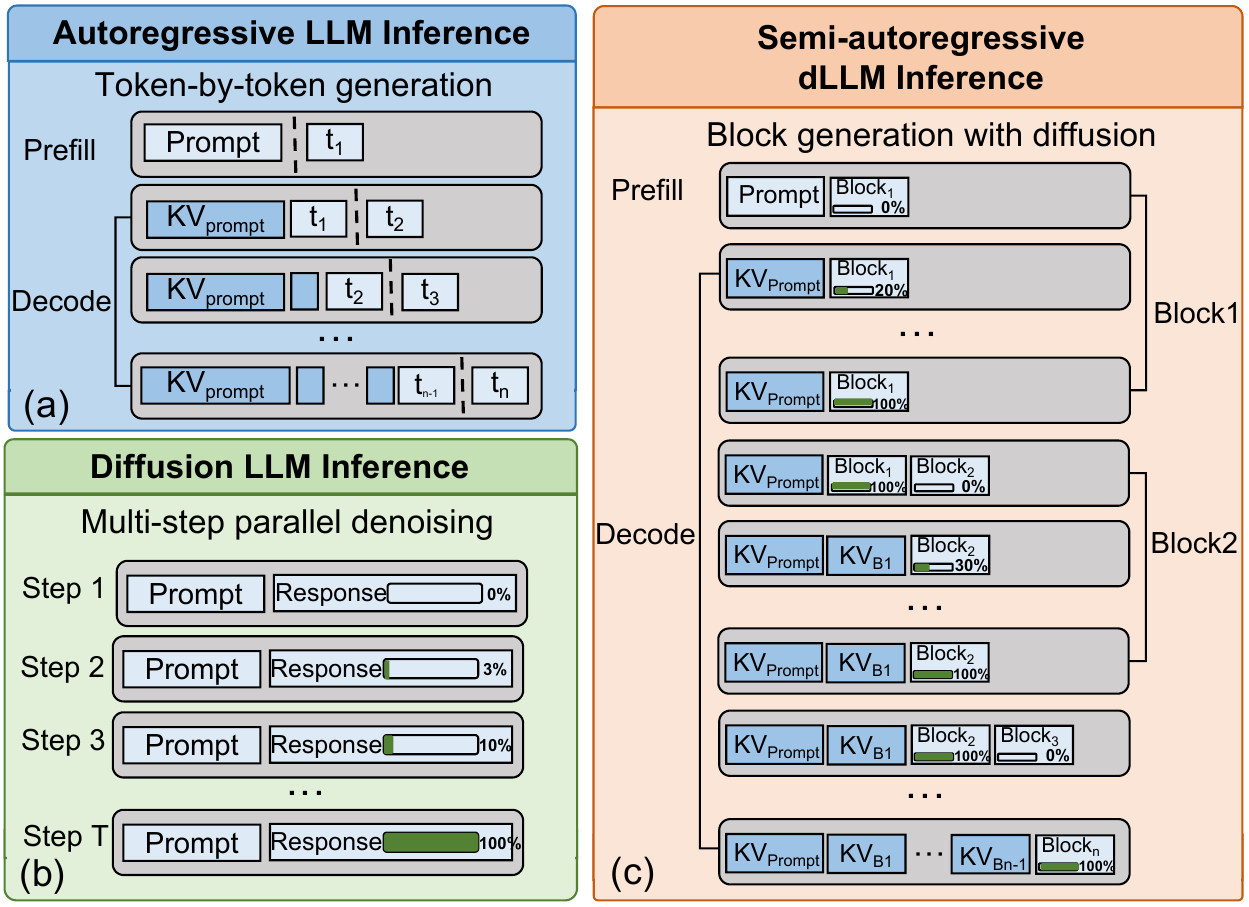}
    \caption{Comparison of three LLM decoding paradigms: (a) autoregressive decoding, (b) fully diffusion decoding, and (c) semi-autoregressive diffusion decoding.}
    \label{fig:Three_Paradigms}
\end{figure}

\parab{Autoregressive MoE-based LLMs.}
Mixture-of-experts (MoE) architectures scale LLM capacity without a proportional increase in inference computation~\cite{deepseekv3,qwen3,glm5,llada2,llada21}. Instead of activating all parameters for every token, MoE models route each token to a sparse subset of expert networks, typically the top-$k$ experts~\cite{shazeer2017outrageously}. Traditional MoE-based LLMs follow autoregressive (AR) generation: after a compute-intensive prefill phase builds the prompt KV cache, decoding generates tokens one by one while reusing cached KV states~\cite{kim2025overfill,yi2026pat}, as shown in \figref{fig:Three_Paradigms}a. Because each decode iteration produces only one token, AR decoding exposes limited token-level parallelism and makes request latency scale with the generated sequence length~\cite{hu2025fast,fu2024break}.

\parab{Fully-diffusion MoE-based dLLMs.}
Diffusion LLMs (dLLMs) have recently emerged as an alternative to AR generation~\cite{llada15,lladav,lladamoe}, with recent variants also adopting MoE layers~\cite{lladamoe}. Instead of generating tokens left to right, fully-diffusion dLLMs perform iterative denoising. As shown in \figref{fig:Three_Paradigms}b, generation starts from a masked response sequence; in each forward pass, or \textit{iteration}, the model predicts all masked positions in parallel, fixes high-confidence tokens, and refines uncertain tokens later~\cite{dllms}. This breaks AR's strict sequential dependency, but repeatedly processing the full sequence introduces substantial redundant computation.

\parab{Semi-AR MoE-based dLLMs.}
Semi-autoregressive (semi-AR) dLLMs balance diffusion parallelism and AR-style efficiency, and they are the primary workload focus of this paper~\cite{llada2,llada21}. As shown in \figref{fig:Three_Paradigms}c, semi-AR dLLMs generate text block by block. Within each block, a diffusion process resolves multiple masked tokens in parallel; across blocks, generation proceeds autoregressively so that KV states of completed blocks can be reused. During prefill, the prompt and the first masked block initialize the KV cache. During decode, the block undergoes denoising iterations until all tokens are finalized; the completed block is committed to the KV cache, and the next masked block is appended.

For consistency, we define an \textit{iteration} as one forward pass through the model. Thus, an AR decode iteration processes one newly generated token, whereas a semi-AR dLLM decode iteration processes a block of active masked tokens.

\subsection{Expert Offloading}\label{subsec:2_2}

While sparse activation reduces per-token computation, the aggregate expert parameter footprint of MoE models exacerbates the GPU memory capacity wall. In recent large-scale MoE models, such as DeepSeek-style, Qwen-style, and LLaDA-style models~\cite{liu2025deepseek,team2026qwen3,llada21}, expert weights account for the dominant fraction of total parameters, often exceeding 90\%. A common way to serve such models under strict GPU memory budgets is \textit{expert offloading}: only a subset of experts is kept in GPU memory, while the rest reside in host CPU memory. We refer to the GPU-resident experts as the \textit{GPU expert cache}. When an activated expert is absent from this cache, the system compensates the miss either by transferring the expert weights to the GPU through host-to-device (H2D) communication or by executing the corresponding expert computation on the CPU~\cite{ktransformers,fiddler,moeinfinity,swapmoe,sida,finemoe}.

\begin{figure}[t]
    \centering
    \includegraphics[width=0.88\columnwidth]{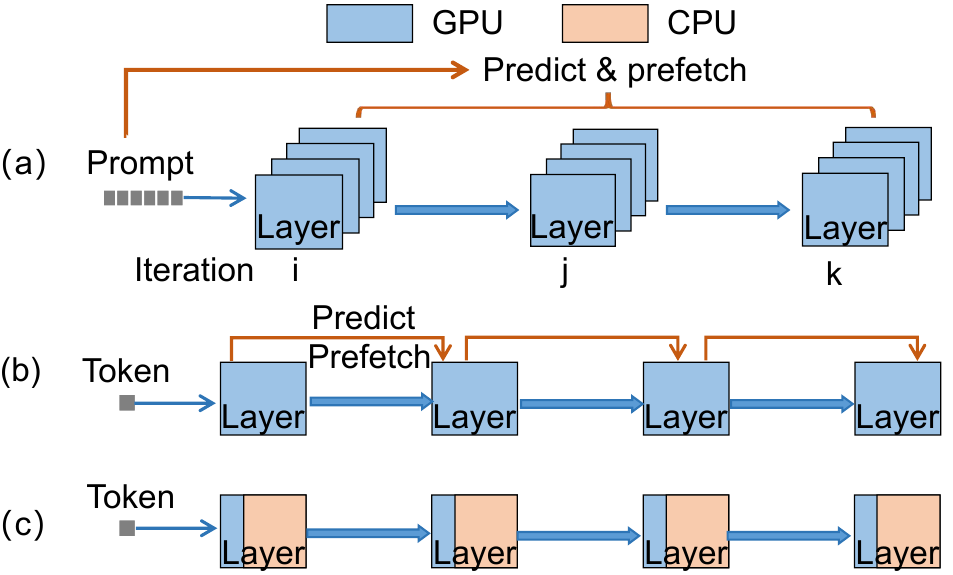}
    \caption{Prior expert offloading paradigms: 
(a) request-level cache, 
(b) intra-iteration prefetch, (c) partial-CPU execution.}
    \label{fig:prior_approach}
\end{figure}

Existing expert offloading systems differ by where expert computation is performed. \textit{All-GPU execution} keeps computation on the GPU and uses H2D transfers for missing experts, while \textit{partial-CPU execution} computes part of the activated experts on the CPU. As shown in \figref{fig:prior_approach}, all-GPU execution further differs by prediction granularity: request-level caching or intra-iteration prefetching.

\parab{All-GPU execution.}
All-GPU systems keep expert computation on the GPU and differ in when they predict expert demand. Systems such as SiDA~\cite{sida} and MoE-Infinity~\cite{moeinfinity} perform request-level caching: they estimate a coarse expert working set for the request and populate the GPU expert cache accordingly (\figref{fig:prior_approach}a). Systems such as SwapMoE~\cite{swapmoe} and FineMoE~\cite{finemoe} perform intra-iteration prefetching: while the GPU computes the current layer, they predict experts needed by subsequent layers and overlap H2D transfers with current-layer computation (\figref{fig:prior_approach}b). These approaches rely on the premise that prediction and transfer latency can be hidden within the current iteration.

\parab{Partial-CPU execution.}
Partial-CPU systems reduce H2D traffic by executing part of the activated expert computation on the CPU. Systems such as KTransformers~\cite{ktransformers} and Fiddler~\cite{fiddler} use a split-expert mechanism: frequently used or performance-critical experts are placed on the GPU, while the remaining activated experts are computed on the CPU (\figref{fig:prior_approach}c). Unlike predictive prefetching systems, these approaches do not rely on explicit expert activation prediction; instead, they exploit idle CPU resources to compensate GPU cache misses when PCIe bandwidth becomes a bottleneck.

\parab{Offloading decision dimensions.}
Prior systems make decisions either before decoding at the request level or within the current decoding iteration, which matches AR decoding where each iteration activates a small expert set. In \secref{subsec:3_1}, we show that semi-AR dLLMs break this assumption and motivate offloading decisions across denoising iterations.
\section{Motivation}\label{sec:3}

This section characterizes semi-AR dLLM workloads (\secref{subsec:3_1}), explains why existing offloading systems fail (\secref{subsec:3_2}), and identifies inter-iteration expert locality as the key opportunity for efficient dLLM offloading (\secref{subsec:3_3}--\secref{subsec:3_4}).

\subsection{Workload Characteristics of MoE-based dLLMs}\label{subsec:3_1}

\begin{figure}[t]
    \centering
    \begin{subfigure}{0.235\textwidth}
        \centering
        \includegraphics[width=\linewidth]{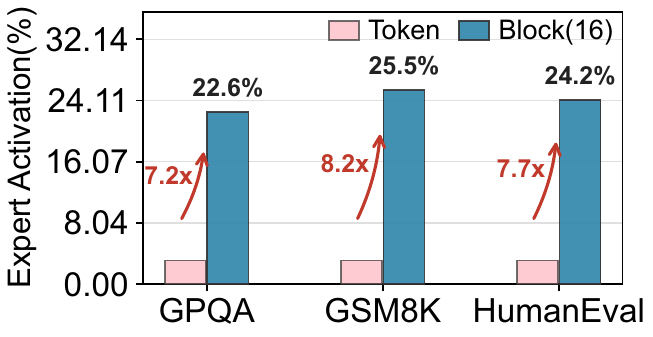}
        \caption{Expert activation footprint}
        \label{fig:expert_activation}
    \end{subfigure}
    \hfill
    \begin{subfigure}{0.235\textwidth}
        \centering
        \includegraphics[width=\linewidth]{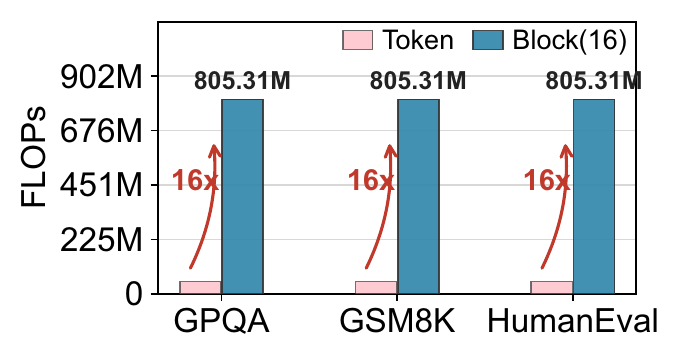}
        \caption{Expert FLOPs}
        \label{fig:expert_flops}
    \end{subfigure}
    \caption{Expert activation footprint and expert FLOPs of semi-AR dLLM decoding compared with AR decoding. Results are measured on LLaDA2.0-mini with block length 16.}
    \label{fig:dllm_workload}
\end{figure}

Block-wise decoding changes the granularity of MoE inference in semi-AR dLLMs. In memory-constrained offloading scenarios dominated by single-request or small-batch decoding, AR decoding routes one generated token per request per iteration, whereas a semi-AR dLLM routes multiple block tokens through MoE layers. This increases both expert activation footprint and expert computation volume.

\parab{Larger expert activation footprint.}
In AR decoding, each MoE layer only needs to access the top-$k$ experts for one newly generated token per request. In contrast, semi-AR dLLMs route multiple active block tokens independently to their respective top-$k$ experts. Although different tokens may share experts, their union remains substantially larger. As shown in \figref{fig:expert_activation}, semi-AR dLLMs activate around $8\times$ more experts per MoE layer per iteration than AR decoding under the same single-request setting.

\parab{Higher expert computation volume.}
Semi-AR dLLMs also increase expert computation because each MoE layer computes expert outputs for all active block tokens rather than one newly generated token. As shown in \figref{fig:expert_flops}, semi-AR dLLMs incur around $16\times$ higher expert FLOPs per MoE layer per iteration than AR decoding.

Overall, semi-AR dLLMs increase GPU cache pressure, H2D transfer demand, and CPU-side compensation cost, directly challenging AR-oriented offloading systems.

\subsection{Limitations of Prior Offloading Systems}\label{subsec:3_2}

To quantify the impact of these workload shifts, we evaluate three expert offloading systems: KTransformers~\cite{ktransformers}, MoE-Infinity~\cite{moeinfinity} and FineMoE~\cite{finemoe}. They cover the paradigms introduced in \secref{subsec:2_2}: request-level GPU caching, intra-iteration GPU prefetching, and partial-CPU execution. We compare them under the same GPU expert-cache budget against two reference points: an ideal cache bound, which maximizes activated-expert residency under the same capacity, and a naive inter-iteration baseline that retains activated experts from the previous iteration. We report TPOT and expert cache utilization, defined as the fraction of GPU expert-cache capacity occupied by experts that are actually accessed.

\begin{figure}[t]
    \centering
    \includegraphics[width=\columnwidth]{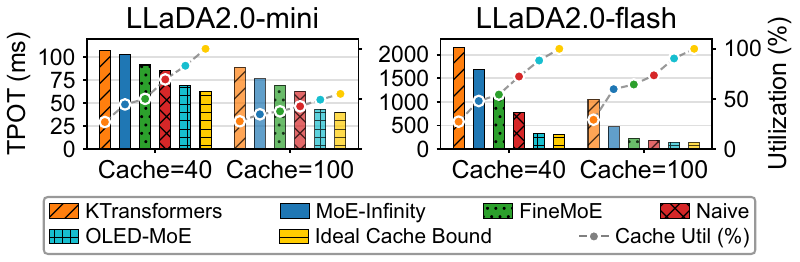}
    \caption{TPOT and expert cache utilization of prior offloading systems under the same GPU expert-cache budget.}
    \label{fig:limit_baseline}
\end{figure}

\begin{figure*}[t]
    \centering
    \begin{subfigure}{0.235\textwidth}
        \centering
        \includegraphics[width=\linewidth]{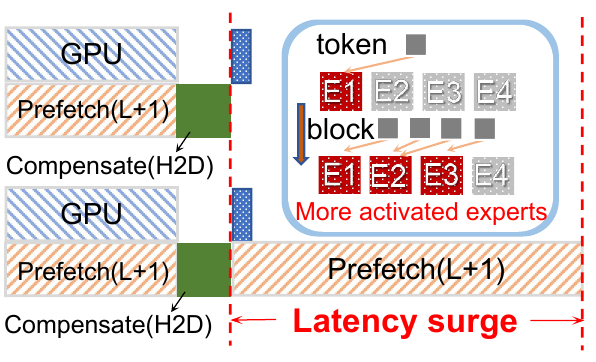}
        \caption{Prefetch failure}
        \label{fig:fail_prefetch}
    \end{subfigure}
    \hfill
    \begin{subfigure}{0.235\textwidth}
        \centering
        \includegraphics[width=\linewidth]{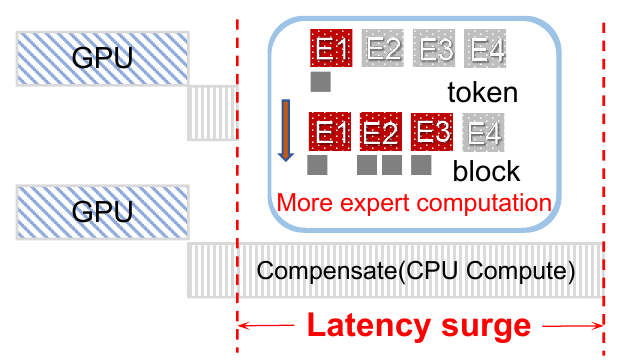}
        \caption{CPU overload}
        \label{fig:fail_cpu_compensation}
    \end{subfigure}
    \hfill
    \begin{subfigure}{0.235\textwidth}
        \centering
        \includegraphics[width=\linewidth]{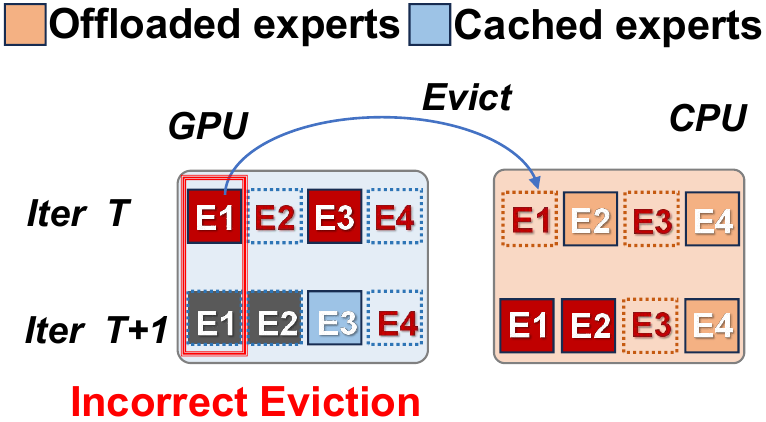}
        \caption{Blind eviction}
        \label{fig:fail_eviction}
    \end{subfigure}
    \hfill
    \begin{subfigure}{0.235\textwidth}
        \centering
        \includegraphics[width=\linewidth]{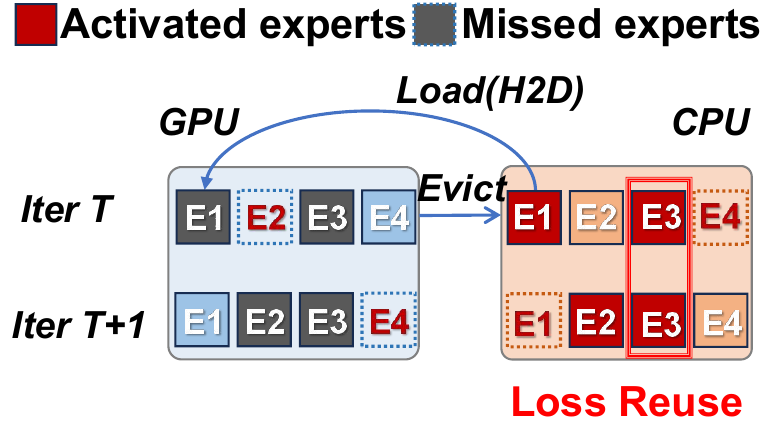}
        \caption{Blind placement}
        \label{fig:fail_placement}
    \end{subfigure}
    \caption{Failure patterns of prior expert offloading systems under MoE-based dLLM workloads: 
(a) exposed prefetch latency, 
(b) exposed CPU compensation latency, 
(c) locality-unaware eviction, and 
(d) reuse-agnostic placement.}
    \label{fig:offloading_failure_patterns}
\end{figure*}

As shown in \figref{fig:limit_baseline}, prior systems remain far from the ideal cache bound under dLLM workloads. MoE-Infinity, FineMoE, and KTransformers achieve TPOTs that are 1.70$\times$--8.16$\times$ the ideal value, while their expert cache utilization is lower by 17.5\%--73.2\%. These results show both expensive miss handling and poor GPU cache utilization. Since dLLM workloads leave little room for intra-iteration recovery, late prefetches and exposed miss compensation make performance depend heavily on the cache state before each layer executes. Thus, eviction or placement decisions that reduce cache utilization become as harmful as slow miss handling. We next analyze this cache-centric bottleneck in all-GPU and partial-CPU systems.

\parab{Reason \#1: all-GPU offloading suffers from exposed prefetch latency and blind eviction.}
All-GPU offloading systems rely on prediction and prefetching to hide H2D transfer latency. This works for AR decoding because each request activates a small expert set, leaving room to transfer later-layer experts during current-layer computation. Under semi-AR dLLM decoding, each MoE layer must serve the aggregated expert demand of an entire block, quickly consuming PCIe bandwidth and leaving insufficient slack for intra-iteration prefetching. As illustrated in \figref{fig:fail_prefetch}, prefetched experts often arrive too late, forcing missing experts onto the critical path.

This exposed prefetch latency makes cache eviction decisions more consequential. LRU/LFU-style policies only reflect past accesses and cannot identify experts useful for future computation. Under dLLM block-wise routing, the enlarged active expert set quickly fills the GPU cache and triggers evictions. As illustrated in \figref{fig:fail_eviction}, useful experts may be evicted by transient current-iteration experts, causing repeated H2D transfers and lower effective cache utilization.

\parab{Reason \#2: partial-CPU offloading suffers from enlarged compensation load and blind placement.}
Partial-CPU systems reduce H2D traffic by executing part of the activated expert computation on the CPU. This approach is effective when cache misses are sparse and CPU-side workload remains small. In semi-AR dLLM inference, block-wise decoding increases expert FLOPs per iteration by an order of magnitude, so each miss can trigger much heavier CPU computation. As illustrated in \figref{fig:fail_cpu_compensation}, the latency of CPU-side expert execution is significantly amplified, substantially prolonging the decoding critical path.

This exposed compensation cost makes CPU/GPU placement decisions more critical. Existing systems make this decision mainly from current H2D and CPU costs. This local cost model may keep useful experts on the CPU while transferring low-value experts to the GPU, reducing future GPU cache utilization. As illustrated in \figref{fig:fail_placement}, reuse-agnostic placement causes repeated CPU execution for valuable experts and lets low-value experts occupy scarce GPU cache.

Overall, these dLLM workloads leave little room for intra-iteration recovery. Once prefetching and compensation fall onto the critical path, offloading performance is dominated by cache utilization before each layer executes. This raises a key question: what information can identify useful experts before they are missed?

\subsection{Insight: Inter-Iteration Expert Locality}\label{subsec:3_3}

The failure analysis above suggests that dLLM offloading needs reuse information beyond the current iteration. Semi-AR dLLMs provide such an opportunity because they refine the same block over multiple denoising iterations. During this process, block hidden states tend to evolve gradually rather than change abruptly. Since MoE gates compute routing decisions from these hidden states, adjacent iterations are likely to preserve many activated experts.

\begin{figure}[t]
    \centering
    \includegraphics[width=0.98\columnwidth]{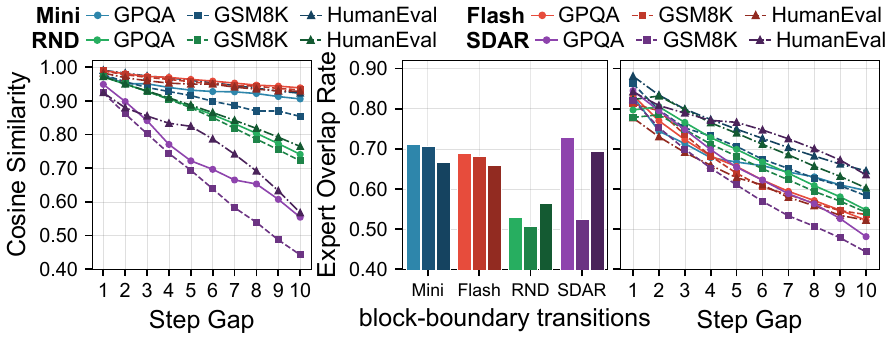}
    \caption{Inter-iteration locality in dLLMs: the left shows hidden-state similarity across iteration gaps, the right shows expert routing overlap across iteration gaps and the middle shows expert routing overlap in block-boundary transitions.}
    \label{fig:insight}
\end{figure}

We quantify this temporal locality using LLaDA2.0-mini and LLaDA2.0-flash on GSM8K~\cite{gsm8k}, HumanEval~\cite{HumanEval}, and GPQA~\cite{gpqa}. As shown in \figref{fig:insight}, hidden states at corresponding layers remain highly similar between adjacent iterations, with cosine similarity reaching 98\%. This continuity translates into stable expert routing: \figref{fig:insight} shows that adjacent iterations share 85\%--90\% of activated experts across models and datasets, and even block-boundary transitions retain more than 65\% overlap. The overlap gradually decreases as the iteration gap grows, indicating a practical temporal window for reuse prediction. Beyond the LLaDA family, SDAR~\cite{sdar} and RND1-Base~\cite{rnd1} exhibit similar trends in hidden-state similarity and expert-routing overlap, suggesting that inter-iteration locality also exists across different dLLM model families.

This observation provides the central insight: dLLM offloading should exploit predictable \textit{inter-iteration} expert locality at the same layer to reserve limited GPU cache for high-value experts, rather than relying on congested \textit{intra-iteration} prediction for subsequent layers.

\subsection{Implications and Challenges}\label{subsec:3_4}

The strong inter-iteration locality identified in \secref{subsec:3_3} suggests that dLLM offloading should exploit same-layer reuse across consecutive denoising iterations. However, turning this locality into an efficient serving system raises two challenges.

\parab{Challenge \#1: predicting reusable experts under limited GPU cache.}
Adjacent iterations exhibit high expert-set overlap, but their active expert sets are not identical. Some experts persist across iterations, while others disappear as token states and routing decisions evolve. Simply retaining all recently accessed experts is therefore insufficient under tight GPU memory budgets: it may preserve transient experts while evicting experts likely to be reused. An effective system must predict which experts remain useful for the same layer in future iterations and translate this prediction into cache-retention decisions.

\parab{Challenge \#2: compensating misses without sacrificing future cache utility.}
Even with accurate prediction, cache misses cannot be eliminated because expert routing changes and GPU memory remains constrained. For each miss, the system must decide whether to fetch the expert to the GPU or execute it on the CPU. This decision should account for both immediate execution cost and future reuse probability: high-reuse experts can amortize transfer cost over future iterations, while low-reuse experts can be served on the CPU to avoid consuming PCIe bandwidth and GPU cache space. Efficient compensation therefore requires jointly optimizing current miss handling and future cache utility.

These challenges motivate \sysname's temporal-locality-aware expert retention and dynamic load- and reuse-aware cooperative compensation in \secref{sec:4}.

\section{Design}\label{sec:4}

\subsection{System Overview}\label{subsec:4_1}

We present \sysname, an inter-iteration locality-aware expert offloading system that converts the temporal locality of MoE-based dLLMs into online cache-retention and miss-compensation decisions. Motivated by the cache-centric bottleneck identified in \secref{subsec:3_2}, \sysname targets one central goal: improving the utilization of the limited GPU expert cache before each MoE layer executes. To this end, \sysname maintains a layer-wise GPU expert cache. Each MoE layer owns an independent cache pool because reusable expert locality in dLLMs primarily appears between the same layer across consecutive iterations. After a layer finishes execution, \sysname updates its cache state to prepare for the same layer in the next iteration.

\begin{figure}[t]
    \centering
    \includegraphics[width=0.9\columnwidth]{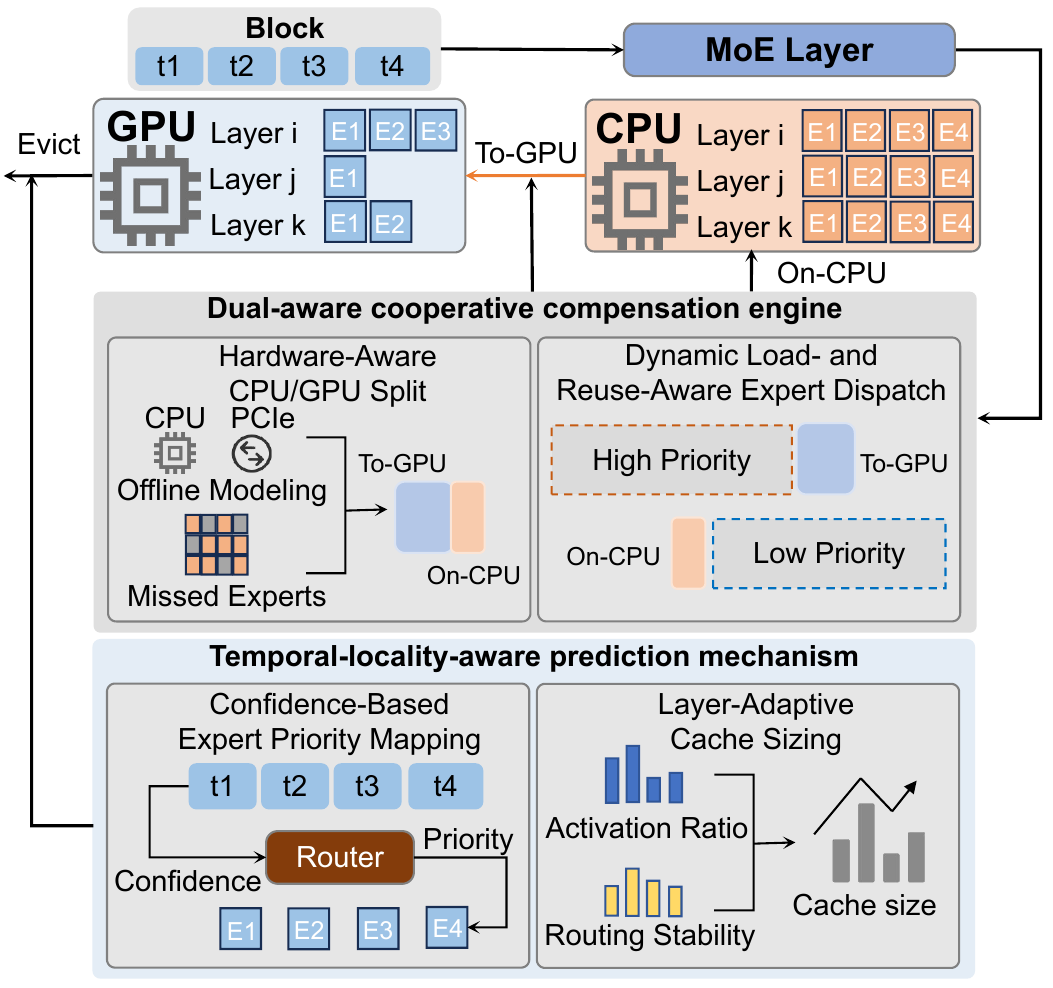}
    \caption{System Overview of \sysname.}
    \label{fig:Design_overview}
\end{figure}

As illustrated in \figref{fig:Design_overview}, \sysname contains two tightly coupled components that operate along the MoE layer execution path. At each layer, the MoE gate first produces the routed expert IDs, gate scores, and token confidence for the current block. The temporal-locality-aware retention mechanism maps these runtime signals to expert-level reuse priorities and adapts each layer's cache capacity according to its activation demand and cache effectiveness. Cache-hit experts execute directly on the GPU, whereas the dynamic load- and reuse-aware compensation engine dispatches cache-missed experts to GPU transfer or CPU execution according to their current token loads and predicted reuse values. After execution, GPU-transferred experts join the cache candidate set, and the retention rule keeps the highest-value candidates under the updated layer budget. The resulting cache state is then reused when the same layer executes in the next denoising iteration. This execution order couples prediction, miss compensation, and cache retention: transferring a missed expert resolves its current computation and can simultaneously warm the cache for future reuse.

\subsection{Problem Formulation}\label{subsec:4_2}

We formulate dLLM expert offloading as an online decision problem that couples current miss-compensation latency with future cache utility. An offline retention plan cannot determine these decisions in advance: the activated expert set depends on the input request and continues to evolve across denoising iterations as token states are refined. Offline profiling can characterize the hardware paths, but it cannot identify which experts are valuable for the current request and iteration. \sysname therefore uses profiling only to establish a hardware-aware cost baseline and makes expert retention, cache allocation, and miss-compensation decisions online from the observed routing state. Consider MoE layer $l$ in denoising iteration $i$. Let $\mathcal{A}_{l}^{i}$ denote the experts activated by the gate, and let $\mathcal{R}_{l}^{i}$ denote the experts resident in the GPU cache of layer $l$ before execution. Experts in $\mathcal{A}_{l}^{i}$ but absent from $\mathcal{R}_{l}^{i}$ become cache misses and require compensation.

For each missed expert, the system must choose one of two compensation paths: transfer it to the GPU and compute it there, or execute it directly on the CPU. Let $\mathcal{M}_{l}^{i}$ denote the missed expert set, and let $\mathcal{G}_{l}^{i}\subseteq\mathcal{M}_{l}^{i}$ denote the missed experts selected for GPU transfer; the remaining experts are executed on the CPU. This decision affects the current layer latency because H2D transfer and CPU execution have different costs. It also affects future cache utility because only GPU-transferred experts can become resident cache candidates for later iterations.

After layer execution, \sysname updates the next-iteration cache state as:
\[
\mathcal{R}_{l}^{i+1}
=
\mathrm{Top}_{C_l^i}
\left(
\mathcal{R}_{l}^{i}
\cup
\mathcal{G}_{l}^{i},
p_{e,l}^{i}
\right),
\]
where $C_l^i$ is the cache capacity of layer $l$, $p_{e,l}^{i}$ is the predicted reuse value of expert $e$, and $\mathrm{Top}_{C_l^i}(\cdot)$ keeps the $C_l^i$ highest-value experts. This rule captures an important coupling: transferring an expert to the GPU is not only a current miss-compensation decision, but also a potential cache warm-up decision for the next iteration.

Ideally, the system should minimize the current compensation latency while preserving experts with high reuse value:
\[
\min_{\mathcal{G}_{l}^{i}\subseteq\mathcal{M}_{l}^{i}}
\quad
T_{\mathrm{comp}}^{i,l}
-
\lambda
\sum_{e\in\mathcal{R}_{l}^{i+1}}
p_{e,l}^{i}.
\]
Here, $T_{\mathrm{comp}}^{i,l}$ is the miss-compensation time of layer $l$, and $\lambda$ conceptually represents the trade-off between current latency and future cache utility rather than a runtime parameter. To solve this problem, \sysname decomposes it into two lightweight online decisions: temporal-locality-aware retention estimates $p_{e,l}^{i}$ and $C_l^i$, while the dynamic load- and reuse-aware compensation engine chooses $\mathcal{G}_{l}^{i}$ using dynamic expert load and predicted reuse value.

\subsection{Temporal-Locality-Aware Expert Retention}\label{subsec:4_3}

\parab{Insight and approach.}
The insight (\secref{subsec:3_3}) suggests that the GPU cache should retain experts for reuse by the same layer in future iterations. However, directly exploiting this locality is non-trivial. The overlap is high but not perfect: some experts persist across iterations, while others are activated only transiently as token states evolve during denoising. Under a limited GPU cache budget, simply retaining recently activated experts, or evicting experts with LRU/LFU, cannot distinguish persistent experts from transient ones. Therefore, \sysname must answer two retention questions: \textit{which} experts are more likely to be reused by the same layer in the next iteration, and \textit{how much} cache capacity each layer should receive. \sysname addresses the first question with confidence- and gate-guided expert priorities, and the second with layer-adaptive cache sizing.

\parab{From token stability to expert priority.}
During each iteration, the model assigns confidence scores to token predictions. These confidence scores are already part of the dLLM decoding algorithm and indicate whether a token prediction is stable enough to be accepted or should continue being refined. This makes token confidence a useful lightweight proxy for routing stability. If a token has high confidence, its hidden state tends to change less in the next denoising iteration, and the MoE gate is more likely to route it to the same experts. Conversely, low-confidence tokens are still unstable and may change their routing decisions.

\begin{figure*}[t]
    \centering
    \begin{subfigure}{0.61\columnwidth}
        \centering
        \includegraphics[width=\linewidth]{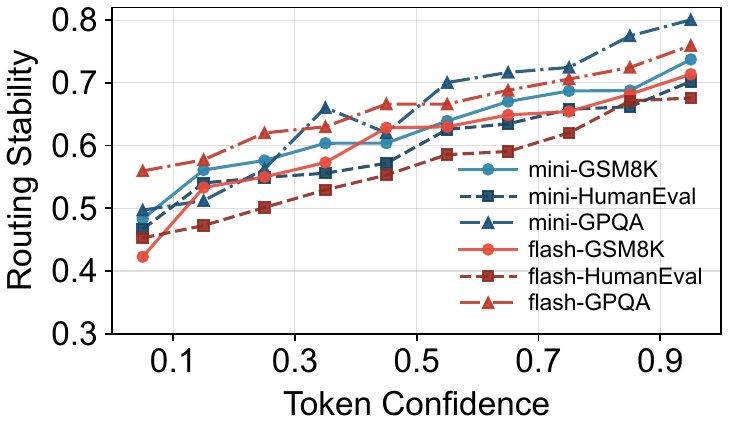}
        \caption{}
        \label{fig:token_confidence}
    \end{subfigure}
    \hfill
    \begin{subfigure}{0.73\columnwidth}
        \centering
        \includegraphics[width=\linewidth]{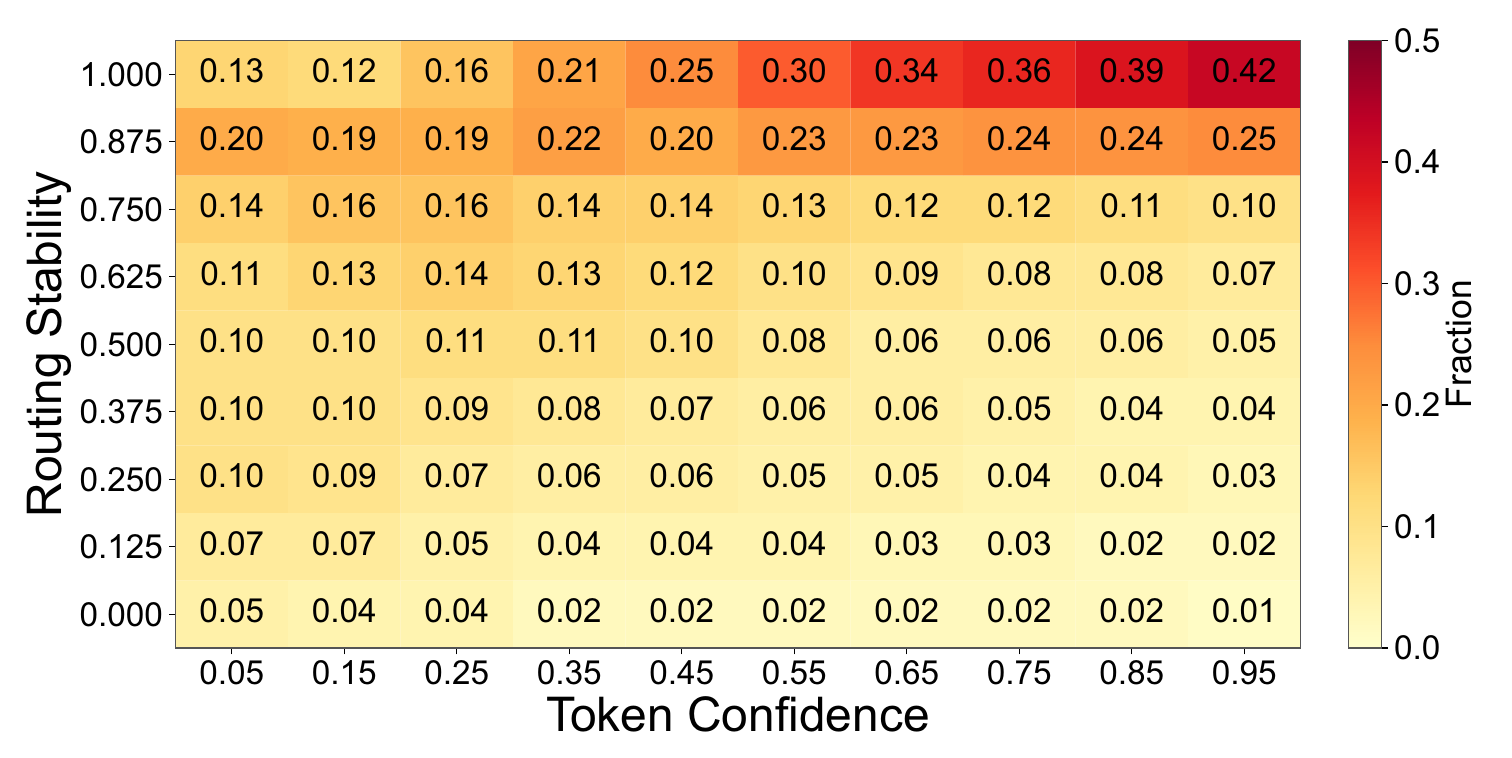}
        \caption{}
        \label{fig:confidence_stability_heatmap}
    \end{subfigure}
    \hfill
    \begin{subfigure}{0.61\columnwidth}
        \centering
        \includegraphics[width=\linewidth]{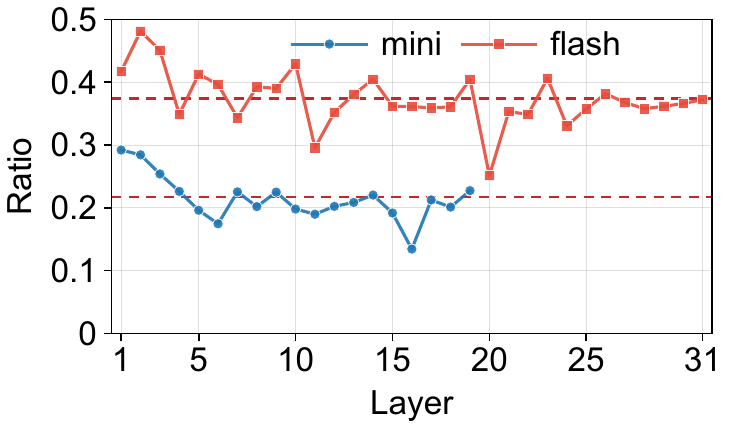}
        \caption{}
        \label{fig:layer_cache}
    \end{subfigure}
    \caption{Design signals used by \sysname: 
(a) token confidence and routing stability, 
(b) confidence-stability distribution, and 
(c) layer-wise activation ratios.}
    \label{fig:design_insight}
\end{figure*}

This finding bridges the gap between the inter-iteration overlap measured in \secref{subsec:3_3} and an actionable cache policy. Instead of treating all activated experts equally, \sysname prioritizes experts selected by stable tokens. As shown in \figref{fig:token_confidence} and \figref{fig:confidence_stability_heatmap}, token confidence is strongly correlated with expert routing stability: high-confidence token groups are more concentrated in high-stability ranges. This provides a lightweight reuse signal without adding a separate predictor or extra model execution.

\begin{figure}[t]
    \centering
    \includegraphics[width=0.95\columnwidth]{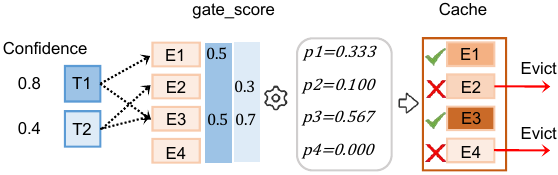}
    \caption{Expert priority mapping from token confidence and gate scores to cache-retention priorities.}
    \label{fig:priority_mapping}
\end{figure}

The current routing observations already capture the information used by conventional cache policies: updating only experts activated in the current iteration reflects recency, while aggregating their gate scores across routed tokens captures activation frequency and routing strength. However, these signals cannot distinguish persistent experts from those activated transiently during denoising. \sysname therefore introduces token confidence as an inter-iteration signal. As shown in \figref{fig:priority_mapping}, token confidence characterizes how likely each token's route is to persist into the next iteration and is used to weight the current routing information, thereby adding temporal stability beyond LRU- and LFU-style policies.

Formally, let $c_t^i$ denote the confidence of token $t$ in iteration $i$, and let $g_{t,e,l}^{i}$ denote the gate score of routing token $t$ to expert $e$ at layer $l$. Let $\mathcal{T}_{e,l}^{i}$ be the set of tokens routed to expert $e$, and let $\mathcal{E}_l$ be the expert set of layer $l$. \sysname computes the expert reuse priority as:
\[
p_{e,l}^{i}
=
\frac{
\sum_{t \in \mathcal{T}_{e,l}^{i}}
g_{t,e,l}^{i} \cdot \psi(c_t^i)
}{
\sum_{e' \in \mathcal{E}_l}
\sum_{t \in \mathcal{T}_{e',l}^{i}}
g_{t,e',l}^{i} \cdot \psi(c_t^i)
},
\]
where $\psi(\cdot)$ is a monotonically increasing function, with $\psi(c)=c$ by default. The numerator aggregates activation frequency and routing strength, while confidence weighting favors routes likely to persist across iterations. The denominator normalizes priorities within each layer and iteration.

After each layer execution, \sysname updates the priorities of experts observed at that layer and evicts low-priority experts if the layer cache exceeds its capacity. For experts that were resident in the GPU cache but are not activated in the current iteration, \sysname sets their priorities to zero. This prevents stale experts from occupying cache space when they are no longer selected by the current denoising trajectory. The priority update uses routing metadata already produced by the MoE gate, so it incurs minimal extra computation.

\parab{From expert priority to layer-adaptive cache sizing.}
The priority rule above uses inter-iteration information to determine which experts are more valuable within each layer, but it assumes that each layer has an appropriate cache budget. Inter-iteration locality also provides a signal at the layer level. As consecutive denoising iterations repeatedly process the same block, the activation demand of each layer evolves relatively stably while remaining heterogeneous across layers. Therefore, the recently observed layer activation state can guide cache allocation for subsequent iterations. As shown in \figref{fig:layer_cache}, LLaDA2.0-mini and LLaDA2.0-flash exhibit clearly different activation ratios across layers. A uniform per-layer partition ignores this persistent layer heterogeneity and may allocate the limited cache capacity to layers that need it less.

This layer heterogeneity becomes more important when applying inter-iteration retention. Different layers vary in both activation demand and the effectiveness of their retained cache entries. Activation ratio alone captures how much cache a layer may need, but does not indicate whether its cached experts are likely to be useful. Conversely, cache utilization alone may favor layers with small activation footprints despite their limited demand. \sysname therefore combines the two signals to allocate cache toward layers that exhibit both high activation demand and effective expert reuse.

\sysname adjusts layer cache sizes using lightweight runtime statistics accumulated over recent denoising iterations. These statistics capture both the activation demand and cache effectiveness of each layer across iterations. Let $a_l$ denote the smoothed activation ratio of layer $l$, and let $u_l$ denote its smoothed expert cache utilization. \sysname computes the layer weight as:
\[
w_l = a_l u_l.
\]
The activation ratio assigns more weight to layers with larger active expert working sets, while cache utilization favors layers whose retained experts are more frequently activated. Their product prevents the allocation from being dominated by either demand or utilization alone.

Given the layer weights, \sysname assigns the total expert-cache capacity through normalized proportional allocation:
\[
C_l =
C_{\mathrm{total}}
\cdot
\frac{w_l}{\sum_j w_j}.
\]
The updated capacity is enforced by the confidence- and gate-guided retention rule after layer execution. In practice, cache sizes are updated every five denoising iterations using smoothed runtime statistics, filtering transient fluctuations while limiting scheduling overhead.

\subsection{Dynamic Load- and Reuse-Aware Cooperative Compensation}\label{subsec:4_4}

\parab{Insight and approach.}
Temporal-locality-aware retention reduces cache misses, but cannot eliminate them because expert routing still changes across denoising iterations and the GPU cache may be smaller than the active expert working set. \sysname must therefore compensate the remaining missed experts through either H2D transfer followed by GPU execution or direct CPU execution. A load-aware split can balance the current CPU and H2D workloads, but it considers only the current iteration. The compensation decision also affects subsequent iterations: a missed expert transferred to the GPU becomes a cache candidate, whereas a CPU-executed expert does not.

To incorporate inter-iteration information into this decision, \sysname reuses the expert priority computed by the retention mechanism as an estimate of future reuse. A missed expert with high reuse priority is more valuable to transfer because it can serve the current iteration and may be reused from the GPU cache in the next iteration. In contrast, a low-priority expert can be executed on the CPU without consuming H2D bandwidth or occupying the cache. \sysname therefore combines the current expert load with the inter-iteration reuse priority: the former balances the immediate CPU and H2D costs, while the latter directs GPU transfers toward experts with greater near-term reuse value.

\parab{Hardware-aware baseline split.}
Before making per-expert decisions, \sysname derives a hardware-aware baseline split from offline profiling. This baseline captures the relative speed of the H2D transfer path and the CPU execution path, preventing the dispatcher from overloading either side.

Let $T_{\mathrm{H2D}}$ denote the profiled latency of transferring one expert from host memory to GPU memory. Let $\tau_{\mathrm{CPU}}$ be the CPU latency of computing one expert for one token, and let $\bar{n}$ be the profiled average number of routed tokens per missed expert. The average CPU compensation cost per missed expert is $\bar{n}\tau_{\mathrm{CPU}}$. \sysname sets the baseline fraction of missed experts assigned to GPU transfer as:
\[
\rho
=
\frac{
\bar{n}\tau_{\mathrm{CPU}}
}{
\bar{n}\tau_{\mathrm{CPU}} + T_{\mathrm{H2D}}
}.
\]
A larger CPU cost increases $\rho$, assigning more missed experts to GPU transfer; a larger H2D cost decreases $\rho$, assigning more missed experts to CPU execution. This ratio is only a starting point: runtime dispatch further adjusts the decision according to each expert's token load and reuse value.

\parab{Dynamic load- and reuse-aware dispatch.}
At runtime, cache-missed experts can differ significantly in both computation load and future value. Let $\mathcal{M}_{l}^{i}$ be the missed expert set and $\mathcal{G}_{l}^{i}$ be the subset selected for GPU transfer. \sysname uses the following objective to guide dispatch:
\[
J(\mathcal{G}_{l}^{i})
=
T_{\mathrm{comp}}^{i,l}
-
\lambda
\sum_{e\in\mathcal{G}_{l}^{i}}
p_{e,l}^{i}.
\]
The first term captures the current-iteration compensation cost using the H2D transfer time, CPU execution time, and token load of each missed expert. The second term introduces inter-iteration information through the reuse priority produced by the retention mechanism. Since only transferred experts can become GPU-cache candidates for subsequent iterations, this term favors transferring missed experts that are more likely to be reused in the near future.

\begin{figure}[t]
    \centering
    \includegraphics[width=1\columnwidth]{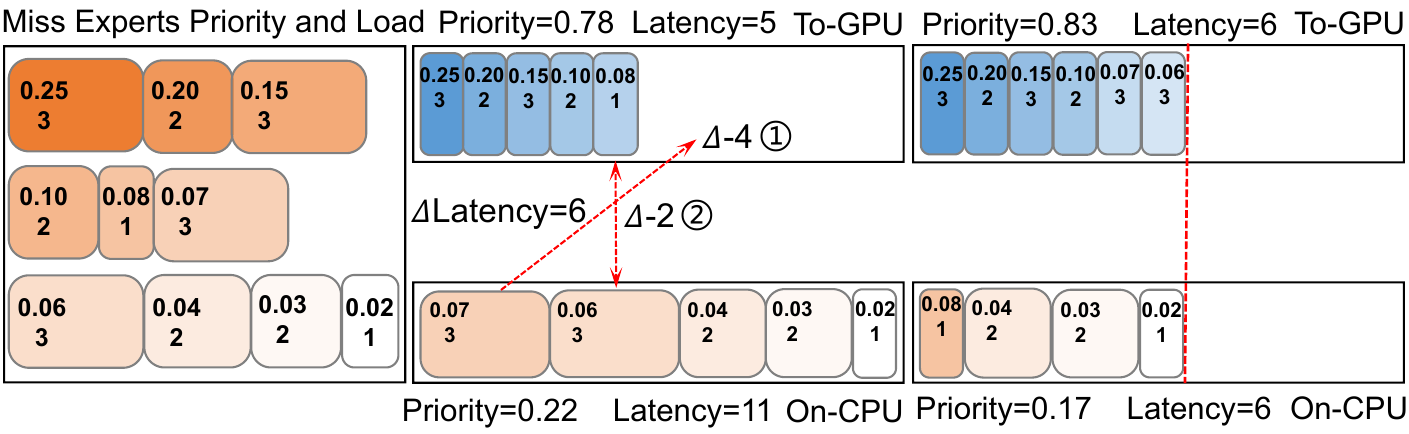}
    \caption{Dynamic load- and reuse-aware dispatch: high-reuse experts are transferred first, then bounded swaps balance CPU/GPU latency.}
    \label{fig:priority_swap}
\end{figure}

Directly minimizing this objective by enumerating all GPU/CPU partitions is too expensive in the decoding loop. As illustrated in \figref{fig:priority_swap}, \sysname therefore uses a two-stage heuristic that separates load balancing from reuse-aware refinement. First, it uses the hardware-aware ratio $\rho$ to bound the number of GPU transfers and selects approximately $\rho|\mathcal{M}_{l}^{i}|$ high-priority experts. This prevents either the CPU path or the H2D path from being overloaded while initializing the GPU-transfer set with experts likely to produce future cache hits. Second, \sysname performs bounded swaps between the GPU-transfer and CPU-execution sets. A swap is accepted only when it improves the estimated objective without moving the partition far from the hardware-aware split. The first stage preserves the current CPU/GPU load balance, while the second improves the future value of the same transfer budget; together, they avoid both load-only placement and a search over all partitions.

\subsection{Implementation}\label{subsec:4_5}

We implement \sysname on top of dInfer v0.1~\cite{dinfer} and use vLLM v0.10.2~\cite{pagedattention} as the generation backend. Instead of modifying the monolithic C++ inference engine, \sysname adopts a minimally intrusive Python-level hook architecture. The hooks are injected into MoE layer definitions to intercept gating outputs, including routed expert IDs, gate scores, and token confidence. These metadata are sufficient for \sysname to update expert priorities, dispatch cache-missed experts, and manage layer-wise cache pools.

\section{Evaluation}\label{sec:5}

\begin{figure*}[t]
    \centering
    \includegraphics[width=\textwidth]{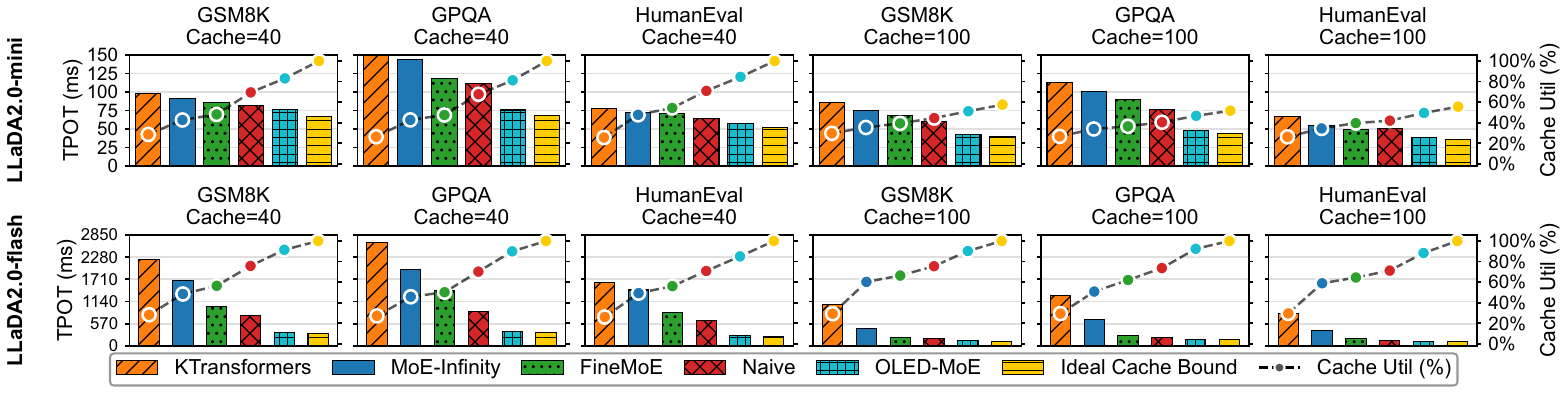}
    \caption{Average TPOT and expert cache utilization of \sysname and baseline systems.}
    \label{fig:overall_tpot_cache}
\end{figure*}

Our evaluation compares \sysname with prior offloading systems in TPOT and expert cache utilization, characterizes its prefill performance through TTFT, analyzes the contribution of each major component, and studies sensitivity to cache capacity, PCIe bandwidth, batch size, and block length.

\subsection{Experimental Setup}\label{subsec:5_1}

\parab{Testbed.}
We evaluate \sysname on two GPU configurations targeting different model scales and memory constraints. Configuration A targets 16B models and uses an NVIDIA GeForce RTX 5090 with 32GB GPU memory. Configuration B targets 100B models and uses an NVIDIA RTX PRO 6000 with 96GB GPU memory. Both use PCIe 5.0 $\times 16$, an Intel Xeon Platinum 8470Q CPU with 25 dedicated cores, CUDA 12.8, and PyTorch 2.8.0. Host memory is 90GB and 240GB for Configurations A and B, respectively.

\begin{figure}[t]
    \centering
    \includegraphics[width=0.49\textwidth]{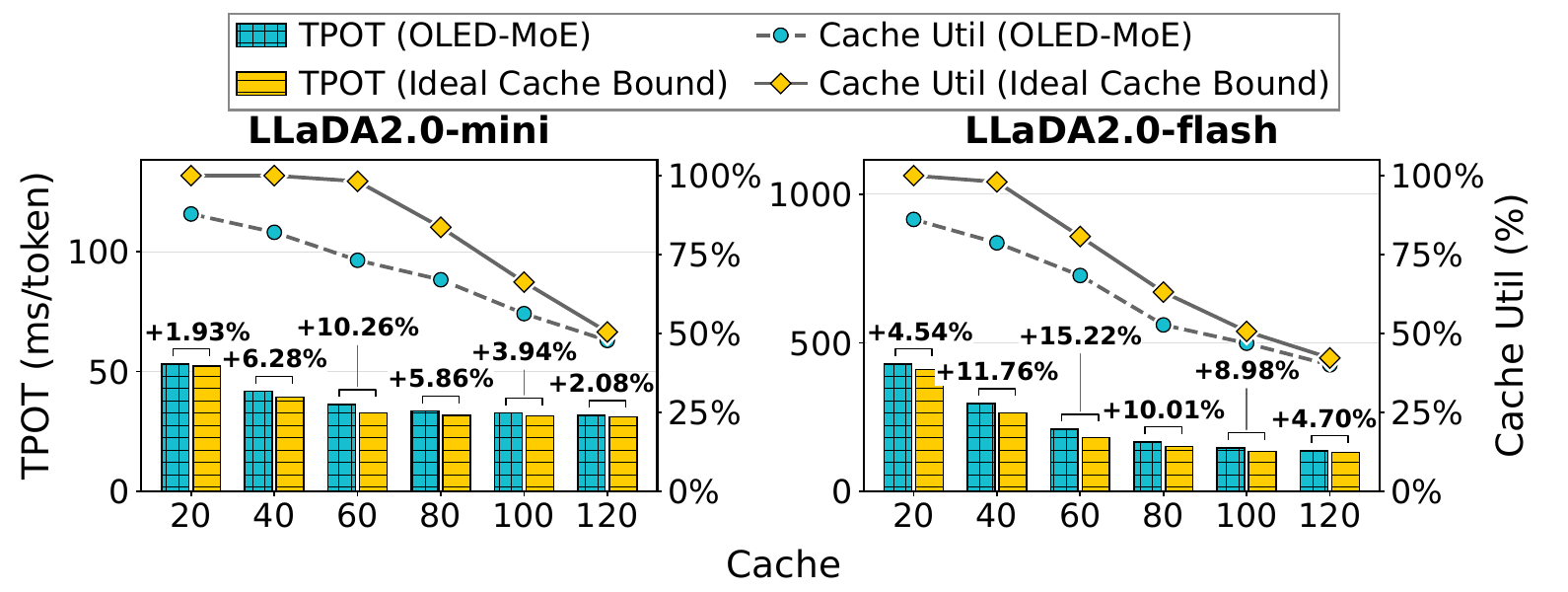}
    \caption{Gap between \sysname and the ideal cache bound across expert-cache capacities.}
    \label{fig:opt_gap}
\end{figure}

\parab{Models and datasets.}
We use two MoE-based semi-AR dLLMs for the main experiments. LLaDA2.0-mini contains 16B parameters, 20 MoE layers, and 256 experts per layer with top-$k$=8. LLaDA2.0-flash contains 100B parameters, 32 MoE layers, and 256 experts per layer with top-$k$=8. Both use block-wise semi-autoregressive generation. We evaluate GSM8K~\cite{gsm8k}, HumanEval~\cite{HumanEval}, and GPQA~\cite{gpqa}. Unless otherwise stated, we use batch size 1, block length 16, and a maximum output length of 1,024 tokens. For mixed-workload experiments, we uniformly sample requests from the three datasets and report the mean. We also report \sysname's TPOT on LLaDA2.1-mini and LLaDA2.1-flash under cache sizes 40 and 100 to test generalization across model versions.

\parab{Baselines.}
We compare \sysname with three representative expert offloading policies. Since existing systems target autoregressive MoE inference and do not directly support semi-autoregressive dLLM execution, we port their core mechanisms into dInfer~\cite{dinfer} using the same backend, GPU expert-cache capacity, CPU worker configuration, and H2D transfer implementation. This isolates the effects of offloading policies from backend differences.

MoE-Infinity~\cite{moeinfinity} represents request-level expert caching, FineMoE~\cite{finemoe} represents intra-iteration later-layer expert prediction and prefetching, and KTransformers~\cite{ktransformers} represents partial-CPU expert execution. We also include Naive-LRU, which maintains an inter-iteration expert cache with LRU replacement, and the ideal cache bound, a clairvoyant upper bound that uses perfect knowledge of the activated experts to maximize the number of activated experts already resident in the GPU cache before each layer executes, subject to the same expert-cache capacity. The ideal cache bound is not a deployable policy and is used only to quantify the best achievable cache effectiveness under the same capacity constraint.

\begin{figure}[t]
    \centering
    \includegraphics[width=0.48\textwidth]{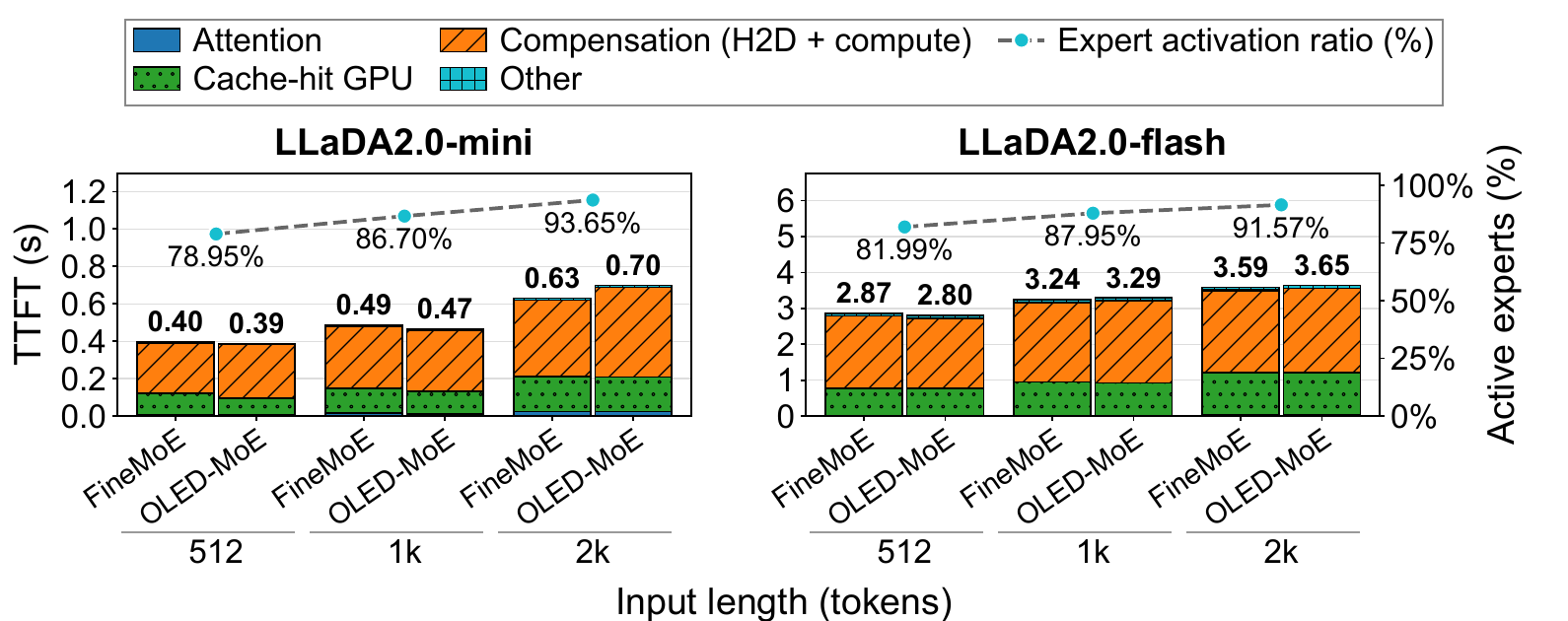}
    \caption{Prefill latency breakdown of \sysname and FineMoE with cache=100.}
    \label{fig:prefill_breakdown}
\end{figure}

\begin{figure*}[t]
    \centering
    \begin{subfigure}{0.48\textwidth}
        \centering
        \includegraphics[width=\linewidth]{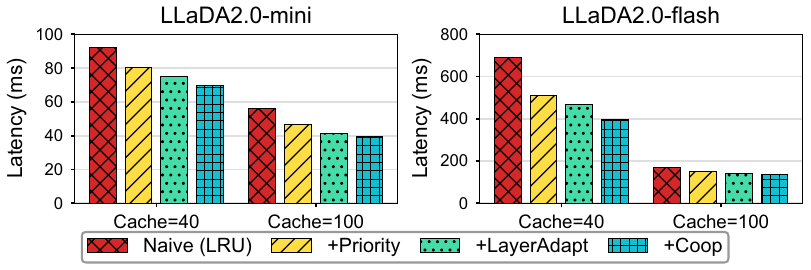}
        \caption{}
        \label{fig:ablation_overall}
    \end{subfigure}
    \hfill
    \begin{subfigure}{0.48\textwidth}
        \centering
        \includegraphics[width=\linewidth]{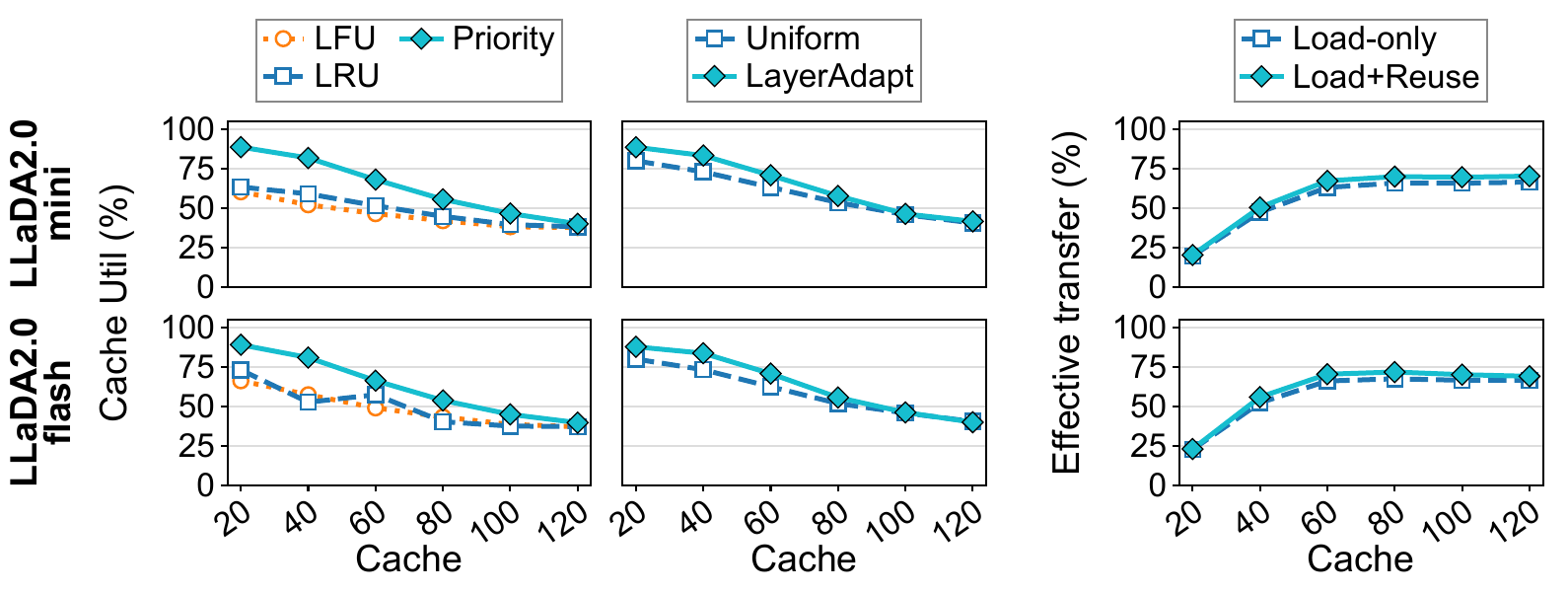}
        \caption{}
        \label{fig:ablation_details}
    \end{subfigure}
    \caption{Ablation analysis of \sysname: (a) incremental TPOT reduction from the three design components; (b) detailed comparisons of components.}
    \label{fig:ablation}
\end{figure*}

\parab{Metrics.}
Our primary latency metric is time per output token (TPOT), measured during decode. Throughout the evaluation, \textit{Cache} denotes the expert-cache capacity, defined as the average number of experts that can reside in GPU memory per MoE layer. We quantify cache effectiveness using expert cache utilization, defined as the fraction of GPU-resident expert-cache slots occupied by experts activated in the corresponding layer execution before miss compensation is triggered. We compute expert cache utilization for each MoE layer and report the average across layers, decoding iterations, and requests. We also report time to first token (TTFT) to characterize prefill performance, although prefill is not the optimization target of \sysname.

\parab{Measurement protocol.}
\sysname is a lossless system optimization: it changes only expert placement and miss handling, without modifying model computation, routing decisions, or generated outputs. Before collecting measurements, we warm up the expert cache once using dummy requests to exclude cold-start effects from the reported results. We repeat each experiment three times and report the mean.

\subsection{Overall Performance}\label{subsec:5_2}

\parab{Overall results.}
We first evaluate the end-to-end decoding performance of \sysname against prior expert offloading baselines. \figref{fig:overall_tpot_cache} reports the average TPOT and expert cache utilization across LLaDA2.0-mini and LLaDA2.0-flash on GSM8K, GPQA, and HumanEval. We evaluate two cache budgets, 40 and 100, denoting the average number of experts cached per layer. Since LLaDA2.0-mini and LLaDA2.0-flash activate about 60 and 90 experts per layer on average, respectively (\figref{fig:layer_cache}), cache size 40 represents a memory-constrained setting, whereas cache size 100 represents a relatively relaxed setting.

Overall, \sysname consistently achieves the lowest TPOT among all practical systems and approaches the ideal cache bound. Compared with the strongest traditional baseline, \sysname reduces TPOT by 1.95$\times$ on average and up to 3.73$\times$ across all settings. This latency improvement is accompanied by higher expert cache utilization: \sysname improves utilization from 51.5\% to 77.7\% on average, corresponding to a 26.2 percentage-point increase over the strongest baseline. These results confirm that \sysname improves decoding performance primarily by making the limited GPU expert cache more useful. By retaining experts that are likely to be reused across denoising iterations, \sysname avoids repeated CPU-GPU expert transfers and reduces miss-induced stalls on the decoding critical path.

The benefit of \sysname is more pronounced when expert misses are more expensive. On LLaDA2.0-flash, \sysname achieves an average TPOT reduction of 4.55$\times$, compared with 1.64$\times$ on LLaDA2.0-mini. This is because each expert miss in the larger model incurs higher transfer and stall cost. The latency variation across datasets mainly stems from differences in the average number of tokens the model can decode per iteration.

\parab{Gap to the ideal cache bound.}
We further compare \sysname with the ideal cache bound defined in \secref{subsec:5_1}. As shown in \figref{fig:opt_gap}, the TPOT of \sysname is only 1.93\%--15.22\% higher than the bound across the evaluated expert-cache capacities. At intermediate capacities, the expert-cache capacity is comparable to the number of activated experts, making performance particularly sensitive to which experts are retained. Consequently, the prediction errors introduced when \sysname estimates future expert reuse from the currently observed routing information and token confidence become more visible. At very small capacities, both systems are severely capacity-constrained, whereas at large capacities, \sysname can retain most reusable experts; the gap therefore narrows in both cases. Overall, \sysname captures most of the performance benefit achievable through expert caching under the same capacity constraint.

\parab{Prefill performance.}
\sysname targets repeated denoising iterations during decode rather than prefill, which occurs only once per request. As shown in \figref{fig:prefill_breakdown}, long prefill sequences activate a large fraction of experts. Consequently, both \sysname and FineMoE naturally achieve high expert cache utilization without specialized cache-management policies. Miss compensation remains the dominant component of TTFT, comprising the H2D transfer and computation of cache-missed experts.

\begin{figure*}[t]
    \centering
    \begin{subfigure}{0.49\textwidth}
        \centering
        \includegraphics[width=\linewidth]{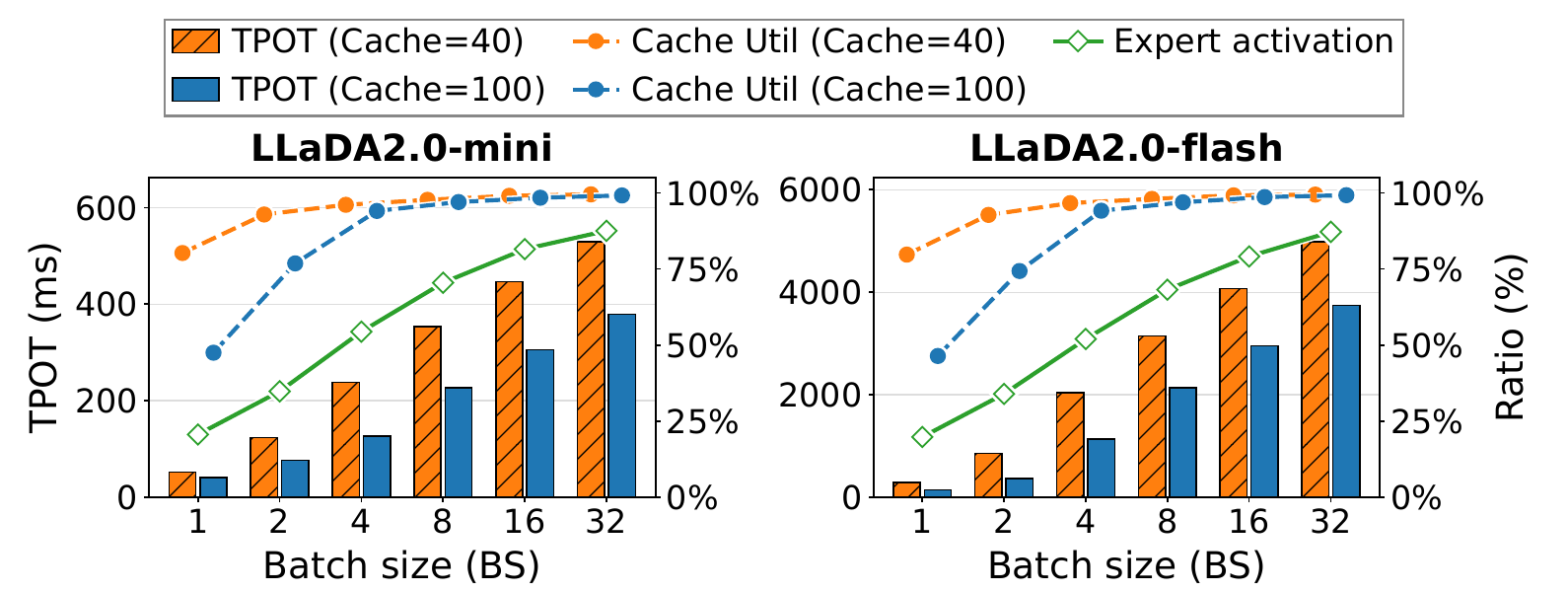}
        \caption{}
        \label{fig:sens_batch}
    \end{subfigure}
    \hfill
    \begin{subfigure}{0.49\textwidth}
        \centering
        \includegraphics[width=\linewidth]{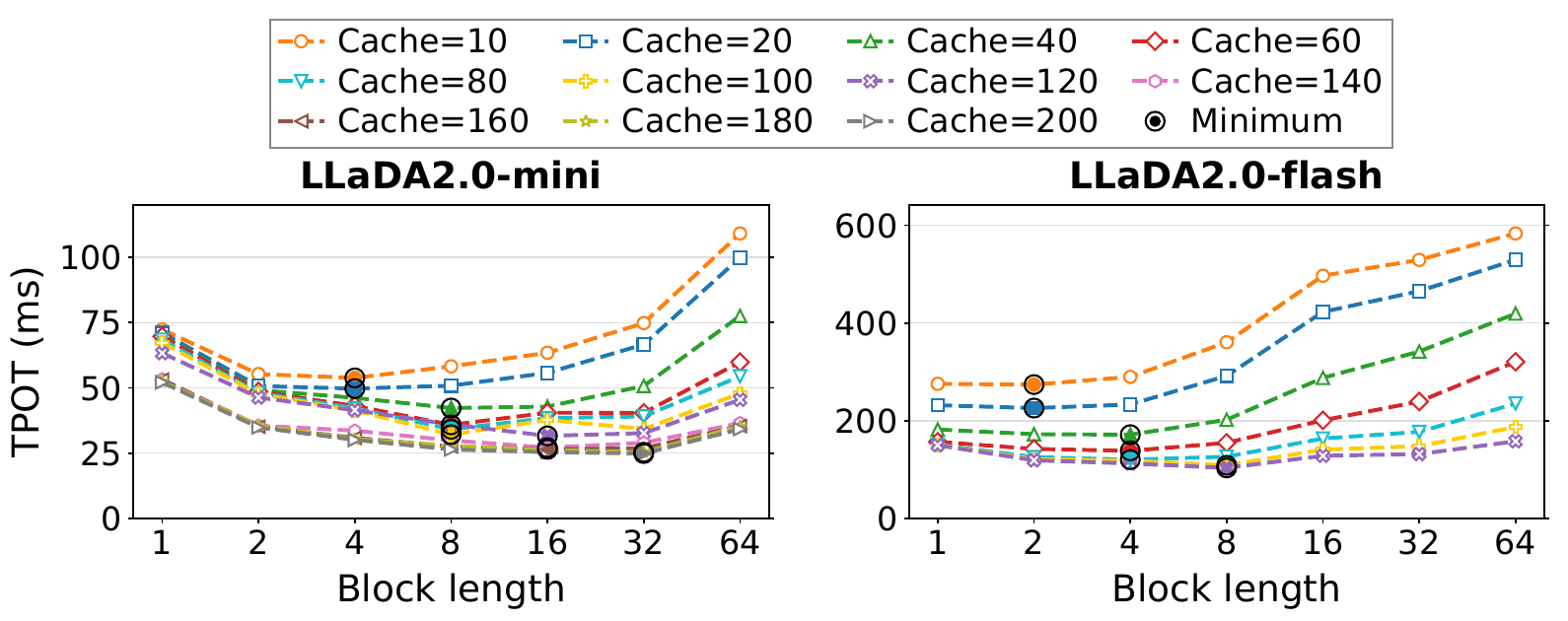}
        \caption{}
        \label{fig:sens_block}
    \end{subfigure}
    \caption{Effects of decoding configurations on expert offloading: (a) TPOT, expert activation ratio, and expert cache utilization of \sysname across batch sizes; (b) TPOT of \sysname across block lengths and expert-cache capacities, with circles marking the minimum TPOT under each capacity.}
    \label{fig:decoding_config}
\end{figure*}

\begin{figure}[t]
    \centering
    \begin{subfigure}{0.235\textwidth}
        \centering
        \includegraphics[width=\linewidth]{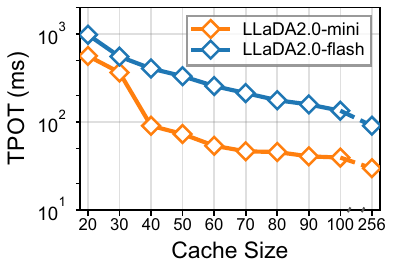}
        \caption{}
        \label{fig:sens_memory}
    \end{subfigure}
    \hfill
    \begin{subfigure}{0.235\textwidth}
        \centering
        \includegraphics[width=\linewidth]{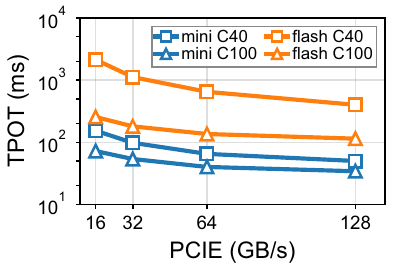}
        \caption{}
        \label{fig:sens_pcie}
    \end{subfigure}
    \caption{Sensitivity of \sysname to (a) GPU expert-cache budget and (b) PCIe bandwidth.}
    \label{fig:sensitivity}
\end{figure}

\subsection{Ablation Study}\label{subsec:5_3}
\parab{Performance breakdown.}
We conduct an ablation study to quantify the contribution of each major component in \sysname. Starting from a simple inter-iteration caching baseline, we incrementally enable the proposed modules and measure the resulting TPOT. The evaluation uses a mixed workload sampled from GSM8K, HumanEval, and GPQA, with the block length fixed to 16. We evaluate LLaDA2.0-mini and LLaDA2.0-flash with average per-layer expert-cache capacities of 40 and 100.

As shown in \figref{fig:ablation_overall}, each component further reduces TPOT. On average, \textit{+Priority}, \textit{+LayerAdapt}, and \textit{+Coop} provide additional TPOT reductions of 20.56\%, 9.09\%, and 9.14\%, respectively. Compared with \textit{Naive-LRU}, the full system reduces TPOT by 53.7\% at Cache=40, versus 34.5\% at Cache=100, and by 50.5\% on LLaDA2.0-flash, versus 37.8\% on LLaDA2.0-mini. These results show that all components contribute to the end-to-end performance improvement, particularly when cache misses are more frequent or more expensive.

\begin{figure}[t]
    \centering
    \begin{subfigure}{0.235\textwidth}
        \centering
        \includegraphics[width=\linewidth]{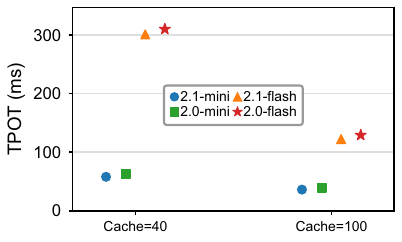}
        \caption{}
        \label{fig:tpot21}
    \end{subfigure}
    \hfill
    \begin{subfigure}{0.235\textwidth}
        \centering
        \includegraphics[width=\linewidth]{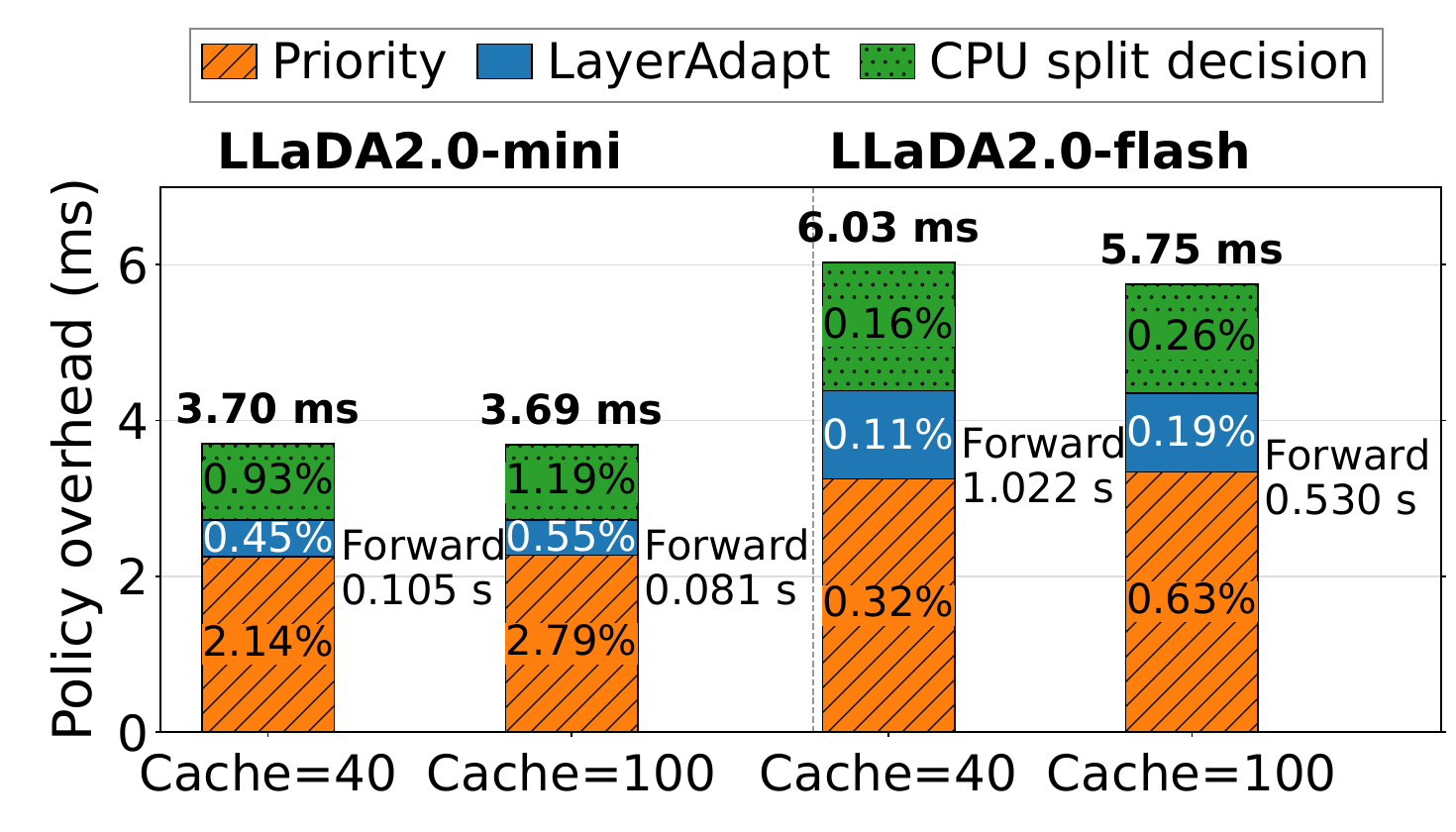}
        \caption{}
        \label{fig:policy_overhead}
    \end{subfigure}
    \caption{Generality and runtime overhead of \sysname: (a) TPOT on LLaDA2.0 and LLaDA2.1 under cache sizes 40 and 100; (b) policy overhead per forward.}
    \label{fig:generality_overhead}
\end{figure}

\parab{Mechanism-level analysis.}
Since Priority and LayerAdapt are designed to improve the effectiveness of the limited expert cache, we further compare their expert cache utilization with alternative policies. For Coop, we evaluate whether reuse-aware dispatch improves the effectiveness of H2D transfers. We define the effective transfer ratio as the fraction of H2D-transferred experts that are retained in the GPU cache and reused in the next denoising iteration.

The first column of \figref{fig:ablation_details} compares Priority with LRU and LFU. At small expert-cache capacities, Priority achieves higher expert cache utilization than LRU and LFU because it weights the current routing information with token confidence, thereby better capturing inter-iteration routing stability and retaining experts that are more likely to be reused. As the cache capacity increases, the cache can retain a broader set of experts. Experts retained without accurate prioritization may also produce incidental cache hits, causing the utilization gap between Priority and the two baselines to gradually narrow.

The second column compares LayerAdapt with uniform cache allocation. LayerAdapt provides the clearest improvement when the average per-layer cache capacity is comparable to the number of activated experts. In this regime, the total cache capacity can cover a substantial portion of the active expert working sets, making demand-aware allocation across heterogeneous layers more effective. When the cache capacity is severely constrained, both policies are subject to similar capacity limitations. When the capacity is relatively ample, both can retain most useful experts. Their expert cache utilization therefore becomes closer in these two regimes.

Finally, the third column compares load- and reuse-aware dispatch with load-only dispatch. Incorporating expert reuse priority consistently increases the effective transfer ratio, indicating that a larger fraction of H2D-transferred experts are retained in the GPU cache and reused in the next denoising iteration. Therefore, reuse-aware dispatch not only balances the current workloads between the CPU and H2D paths, but also directs the limited transfer budget toward experts with greater near-term reuse value.

\subsection{Sensitivity and Overhead}\label{subsec:5_4}

\parab{Scaling with batch size.}
We evaluate \sysname across batch sizes from 1 to 32 under expert-cache capacities of 40 and 100 experts per layer. As shown in \figref{fig:sens_batch}, increasing the batch size increases the number of concurrently processed tokens and enlarges the union of activated experts. Consequently, the expert activation ratio increases and the expert-level sparsity of the MoE model gradually decreases. Meanwhile, expert cache utilization approaches saturation because a large fraction of the cached experts are activated at larger batch sizes. Although a larger cache capacity continues to reduce TPOT, the increasing expert activation footprint narrows the headroom for expert-cache management. These results show that expert offloading is most effective for single-request and small-batch serving, where the active expert working set remains sufficiently sparse for selective cache retention.

\parab{Cache-dependent block length.}
Block length creates a trade-off between decoding parallelism and expert-offloading efficiency. A larger block processes more tokens per forward pass, but it also activates a larger expert working set and increases pressure on the limited GPU expert cache. We evaluate block lengths from 1 to 64 across a broad range of cache capacities. Block length 1 provides an AR-style token-by-token execution reference under the same model, hardware, and memory constraints, while larger block lengths progressively expose the parallelism of semi-autoregressive decoding. As shown in \figref{fig:sens_block}, block length 1 incurs higher TPOT than the best semi-autoregressive configuration across the evaluated cache capacities, but the TPOT-minimizing block length shifts with the available cache. Smaller blocks are preferred under tight cache budgets, whereas larger caches can support longer blocks before their expanded expert working sets dominate offloading cost. These results show that semi-autoregressive execution can outperform the AR-style reference under the same system constraints, while its block length should be selected jointly with the available expert-cache capacity.

\parab{Impact of cache capacity.}
We evaluate \sysname under different expert-cache capacities on LLaDA2.0-mini and LLaDA2.0-flash, and include Cache=256 as the full-residency bound. As shown in \figref{fig:sens_memory}, TPOT decreases as the expert-cache capacity increases. The reduction is particularly pronounced as the cache capacity approaches the number of activated experts, because each additional cache slot can retain an expert that would otherwise require miss compensation. Once most activated experts can reside on the GPU, further increases in capacity provide diminishing returns. At Cache=100, \sysname uses only about 40\% of the expert GPU memory required for full expert residency, while its TPOT is only 32\% and 23\% higher than the full-residency bound for LLaDA2.0-mini and LLaDA2.0-flash, respectively. The full-residency result for LLaDA2.0-flash is estimated because its complete expert set cannot fit on a single GPU.

\parab{Impact of PCIe bandwidth.}
We emulate effective PCIe bandwidths from 16 to 128\,GB/s by scaling the H2D transfer latency inversely with the target bandwidth, and evaluate both models with Cache=40 and Cache=100. As shown in \figref{fig:sens_pcie}, TPOT decreases as the effective PCIe bandwidth increases, confirming that PCIe bandwidth is an important performance factor in expert-offloading scenarios. The reduction is particularly pronounced under Cache=40, where the limited cache capacity results in more cache misses and a larger fraction of experts being transferred from the CPU to the GPU. Under Cache=100, more activated experts remain resident on the GPU, reducing H2D traffic and making TPOT less sensitive to PCIe bandwidth.

\parab{Generalization across model versions.}
We additionally evaluate \sysname on LLaDA2.1-mini and LLaDA2.1-flash under Cache=40 and Cache=100, using the same experimental settings as for LLaDA2.0. As shown in \figref{fig:tpot21}, increasing the expert-cache capacity consistently reduces TPOT across both LLaDA2.0 and LLaDA2.1, and the corresponding model variants exhibit comparable performance trends. Although this experiment does not include a full baseline comparison, the consistent behavior indicates that \sysname is not specific to a single LLaDA release. This is consistent with our observation in \secref{subsec:3_3} that inter-iteration expert locality arises from the iterative denoising process of semi-autoregressive dLLMs.

\parab{Runtime policy overhead.}
We measure the runtime cost of expert-priority computation, layer-adaptive cache allocation, and CPU/GPU split decisions. As shown in \figref{fig:policy_overhead}, these policies together add 3.69--6.03\,ms per full-model forward, corresponding to 0.59\%--4.53\% of the forward latency across the evaluated models and expert-cache capacities. Expert-priority computation is the largest component, while layer adaptation and CPU/GPU split decisions contribute smaller fractions. The relative overhead is lower on LLaDA2.0-flash because its model forward is substantially longer. Overall, the online policies introduce limited runtime cost compared with the model computation they guide.

\section{Related Work}\label{sec:6}

\parab{Expert offloading for MoE inference.}
Expert offloading has been widely studied for serving large-scale MoE LLMs under limited GPU memory. Existing systems improve offloading efficiency through predictive caching, expert prefetching, routing-aware execution, or fine-grained expert management~\cite{he2024expertflow,chen2026firm,li2026commitmoe,zhao2026mobile,lin2025depth,wang2025od}. These techniques are primarily designed for autoregressive (AR) decoding, where each iteration activates a small expert set and intra-iteration prediction remains effective. In contrast, \sysname targets MoE-based dLLMs, whose block-wise decoding creates large concurrent expert activations and makes AR-oriented prefetching ineffective. \sysname exploits inter-iteration expert locality to predict and retain reusable experts across denoising iterations.

\parab{CPU-GPU hybrid inference for MoE models.}
CPU-GPU hybrid MoE systems alleviate GPU memory pressure by combining GPU-side expert caching, CPU-side cache-miss handling, expert placement, and overlapped transfer and computation~\cite{zhu2026dali,zhong2025hybrimoe,zhang2025daop,liu2025remoe,huang2026efficient,zhang2025duoserve}. These systems mainly optimize CPU/GPU placement according to current execution or transfer cost. They do not account for the multi-iteration expert reuse pattern of semi-AR dLLM decoding, where transferring a missed expert to the GPU can also improve future cache efficiency. \sysname therefore introduces a dynamic load- and reuse-aware cooperative compensation engine that jointly considers current CPU-GPU load balance and future inter-iteration reuse probability.

\parab{Diffusion LLM inference acceleration.}
Recent work accelerates diffusion-based LLM inference through block diffusion, hierarchical caching, confidence-aware calibration, dynamic block or step control, pruning, early exiting, and speculative decoding~\cite{wu2025fast,shen2025improving,chen2026dflash,xiao2026treaming,agrawal2025spiffy,sandler2025specdiff,zhengmosaic}. These techniques reduce decoding computation or latency, but they do not address the memory pressure caused by large MoE expert weights. To our knowledge, \sysname is the first expert offloading system designed specifically for MoE-based dLLMs, exploiting inter-iteration expert locality to improve offloading efficiency.

\parab{Orthogonal memory-saving techniques.}
KV cache offloading and weight quantization are complementary to expert offloading. KV cache paging or offloading reduces attention-state memory~\cite{nian2026cacheflowefficientllmserving,sun2024shadowkv,cho2024kv,zhang2023h2o,tang2024quest,hu2025tightllm}, while low-bit quantization reduces expert weight size and H2D transfer volume~\cite{liu2024kivi,hooper2024kvquant}. \sysname focuses on expert-weight offloading and can be combined with these techniques through a unified memory manager that balances KV cache capacity, expert cache capacity, and quantization-induced compute overhead.

\section{Conclusion}\label{sec:8}

We presented \sysname, an expert offloading system for MoE-based dLLM inference under constrained GPU memory. \sysname addresses the mismatch between AR-oriented offloading and semi-AR dLLM execution by shifting the optimization target from intra-iteration prefetching to inter-iteration expert retention. To improve GPU expert cache utilization, \sysname combines temporal-locality-aware expert retention, layer-adaptive cache allocation, and dynamic load- and reuse-aware cooperative compensation. Evaluation results show that \sysname reduces TPOT by 1.23$\times$--7.93$\times$ and improves expert cache utilization by 1.44$\times$--4.23$\times$ over state-of-the-art baselines, demonstrating that inter-iteration expert locality is an effective scheduling dimension for memory-constrained MoE-based dLLM serving.
\section*{Acknowledgments}

We thank the anonymous reviewers and our shepherd for their insightful feedback. This work was supported by the National Natural Science Foundation of China under Grant No.~62572341.

\bibliographystyle{ACM-Reference-Format}
\bibliography{refer}
\end{document}